\documentclass[aps,prl,twocolumn,superscriptaddress,longbibliography]{revtex4-2}%
\usepackage{graphicx}
\usepackage{float}
\usepackage{dcolumn}
\usepackage{bm,color}
\usepackage{amsmath,amssymb,dsfont,amstext,amsfonts}
\usepackage{extarrows}
\usepackage{xcolor}
\usepackage{wasysym}
\usepackage{mathtools}
\usepackage{bbold}
\usepackage{hyperref}

\hypersetup{
breaklinks=true,
colorlinks=true,
citecolor=blue,
linkcolor=black,
filecolor=black,
urlcolor=blue
}
\usepackage[normalem]{ulem} 
\usepackage{color}%makes \textcolor{grey}{text} available

\usepackage{titlesec}
\titleformat{\section}
  {\centering\normalfont\fontsize{12}{15}\bfseries}{\thesection}{1em}{}
\makeatletter
\def\@hangfrom@section#1#2#3{\@hangfrom{#1#2}#3}%\MakeTextUppercase{#3}}%
\def\@hangfroms@section#1#2{#1#2}%\MakeTextUppercase{#2}}%
\makeatother

\def\bra#1{\mathinner{\langle{#1}|}}
\def\ket#1{\mathinner{|{#1}\rangle}}

\def\braket#1{\mathinner{\langle{#1}\rangle}}

\def\dd{\mathrm{d}}
\newcommand{\mr}[1]{\mathrm{#1}}
\newcommand{\dg}{\dagger}
\newcommand{\mb}{\mathbf}

\begin{document}

\title{Exact projected entangled pair ground states with topological Euler invariant}
%\title{Exact topological PEPS networks inspired by ideal quantum geometry}
%\author{tw and people and rjs}
\author{Thorsten B. Wahl}
\email{tw344@cam.ac.uk}
\affiliation{TCM Group, Cavendish Laboratory, Department of Physics, J J Thomson Avenue, Cambridge CB3 0HE, United Kingdom}

\author{Wojciech J. Jankowski}
\affiliation{TCM Group, Cavendish Laboratory, Department of Physics, J J Thomson Avenue, Cambridge CB3 0HE, United Kingdom}

\author{Adrien Bouhon}
\affiliation{TCM Group, Cavendish Laboratory, Department of Physics, J J Thomson Avenue, Cambridge CB3 0HE, United Kingdom}
\affiliation{Nordita, Stockholm University and KTH Royal Institute of Technology, Hannes Alfv{\'e}ns v{\"a}g 12, SE-106 91 Stockholm, Sweden}

\author{Gaurav Chaudhary}
\affiliation{TCM Group, Cavendish Laboratory, Department of Physics, J J Thomson Avenue, Cambridge CB3 0HE, United Kingdom}

\author{Robert-Jan Slager}
\email{rjs269@cam.ac.uk}
\affiliation{TCM Group, Cavendish Laboratory, Department of Physics, J J Thomson Avenue, Cambridge CB3 0HE, United Kingdom}
\date{\today}

\begin{abstract}
We report on a class of gapped projected entangled pair states (PEPS) with non-trivial Euler topology motivated by recent progress in band geometry. 
In the non-interacting limit, these systems have optimal conditions relating to saturation of quantum geometrical bounds, allowing for parent Hamiltonians whose lowest bands are completely flat and which have the PEPS as unique ground states. Protected by crystalline symmetries, these states evade restrictions on capturing tenfold-way topological features with gapped PEPS. These PEPS thus form the first tensor network representative of a non-interacting, gapped two-dimensional topological phase, similar to the Kitaev chain in one dimension. 
Using unitary circuits, we then formulate interacting variants of these PEPS and corresponding gapped parent Hamiltonians. We reveal characteristic entanglement features shared between the free-fermionic and interacting states with Euler topology. Our results hence provide a rich platform of PEPS models that have, unexpectedly, a finite topological invariant, forming the basis for new spin liquids, quantum Hall physics, and quantum information pursuits. 
% {\color{green} In "Euler class" we're saying that the Euler insulators (and the PEPS) are Wannierrisable if $C_2T$ symmetry is not enforced}
\end{abstract}

\maketitle
\section{Introduction} 

Tensor network states (TNS) form a generally applicable tool for the description of quantum matter. A numerically efficient representation of the ground states of local gapped Hamiltonians~\cite{Verstraete2006,Huang2014,Molnar2015,Dalzell2019,Huang2019}, TNS play a pivotal role both in the simulation of correlated systems~\cite{White1998,Yan2011,Corboz2013,Corboz2014,He2017,Gohlke2018} and the analytical classification of topological phases~\cite{Pollmann2010,Schuch2011,Williamson2016,Wahl2018,Chan2020,Li2020}. 
%TNS thus form a generally applicable tool for the description of quantum many-body systems. 
Yet, due to the increased complexity in higher dimensions, TNS have not yet matched the success of non-interacting band theory~\cite{Rmp1, Rmp2}, both in simulations and the classification of topological phases~\cite{Clas1, Clas2, clas4,  Clas5, Clas6, Clas7, Shiozaki14}. Of particular interest are therefore systems which can be well captured with band theory but are marked by difficulties when it comes to TNS approaches. The most well-known such example is chiral topological systems: In topological band theory, they are characterized by occupied bands whose overall Chern number is non-vanishing, separated by a gap from the conduction bands. While TNS approaches are able to capture the chiral topological features of such systems, this comes at the cost of producing algebraically decaying correlations characteristic of critical systems~\cite{Dubail2015,Wahl2013,Yang2015}. Generally, it has been shown that TNS with exponentially decaying correlations cannot capture any higher-dimensional topological invariant~\cite{Kitaev} of the ten Altland-Zirnbauer (AZ) classes~\cite{Read2017}. 

The severe restrictions of TNS to represent gapped non-interacting topological phases suggest that TNS might equally struggle to capture topological phases protected by crystalline symmetries. However, 
%An important question is whether the above limitations of TNS extend to the case of topological phases with crystalline symmetries. 
%DMRG, tensor networks and PEPS solutions provide a rich playground to study correlated systems and topological ones in particular \textcolor{orange}{ @thorsten please add amend this with a few sentences and references of mile stones etc}. While topological band theory~\cite{Rmp1, Rmp2} has rapidly progressed over the past decades~\cite{Clas1, Clas2, clas4,  Clas5, Clas6, Clas7, Shiozaki14}, it is a priori not evident that these advances can be translated to these contexts. 
reinvigorated interests~\cite{tormaessay, bouhon2023quantum} in relation to quantum geometry~\cite{provost1980riemannian, resta_2011_metric} could provide a useful tool in that they outline flatband conditions. Under such conditions, %interactions take a more forefront role, but, more importantly, 
it is possible to define topological flatband Hamiltonians that can be formulated as exemplary parent Hamiltonians. These are sums of projectors with local support that each annihilate the ground state(s). From these local projectors, a TNS ground state can in principle be constructed. However, whether the resulting state is non-vanishing and can be made the unique ground state of such a crystalline symmetry-protected topological Hamiltonian might be hampered by the previously mentioned hurdles. Intuitively, TNS with exponentially decaying correlations are incompatible with topological invariants, as they come with delocalized edge modes around a physical boundary (in more than one dimension); the local structure of tensor networks is incapable of separating such edge modes from the bulk modes, delocalizing them as well. Because of that, in the following, we consider crystalline symmetry-protected topological phases which do not have helical or chiral edge modes. 

We show that a family of topological projected entangled pair states (PEPS)~\cite{PEPS} and generating Hamiltonians can be formulated in the context of the Euler class~\cite{bouhon2019nonabelian, Clas7, BJY_nielsen, Bouhon2018Wilson}. The Euler class is a multi-gap invariant~\cite{Clas7}, pertaining to topological structures that emerge when groups of partitioned bands (band subspaces) carry non-trivial topological indices~\cite{Clas7}. % as set by homotopy charges of the defining Grassmannian or flag manifold~\cite{Clas7}. 
These topological charges of groups of bands can be altered by braiding nodes in momentum space, as band nodes residing between neighboring bands can carry non-Abelian charges~\cite{bouhon2019nonabelian, BJY_nielsen, doi:10.1126/science.aau8740}.
The braiding of non-Abelian frame charges and multi-gap topologies have been increasingly related, both theoretically and experimentally, to physical systems that range from  out-of-equilibrium settings~\cite{slager2024floquet, Unal_2020, Zhao_2022} and 
phonon spectra~\cite{Peng2021, Peng2022Multi}, to electronic systems (twisted, magnetic and conventional)~\cite{magnetic, BJY_nielsen, lee2024eulerbandtopologyspinorbit} as well as     metamaterials~\cite{Jiang_2021, JIANG2024, Guo1Dexp}.

The here introduced family of PEPS has non-trivial Euler class, evading no-go conditions as for tenfold-way topologies~\cite{Read2017}, constituting a 2D %(albeit non-Wannierizable) 
analogue of the Kitaev chain~\cite{Kitaev2001}. %, due to an elusive bulk-boundary correspondence.
These PEPS differ from the higher-order topological insulators represented by
PEPS in Ref.~\cite{Hackenbroich2020}, which have a bond dimension in one spatial direction that grows with the system size.
Within the non-interacting limit, from a band theory perspective, our PEPS enjoy ideal quantum geometrical properties, which we elucidate in the Supplementary Information (SI) in detail. More importantly, by applying shallow quantum circuits of diagonal
unitaries, we transform these PEPS and their gapped parent Hamiltonians to interacting variants. We can signify the Euler phase both in the non-interacting and the interacting limit upon appealing to the entanglement spectrum. As such, our results set a benchmark for an exact class of PEPS parent Hamiltonians with  finite topological invariant.

\begin{figure}[t]
\includegraphics[width=0.45\textwidth]{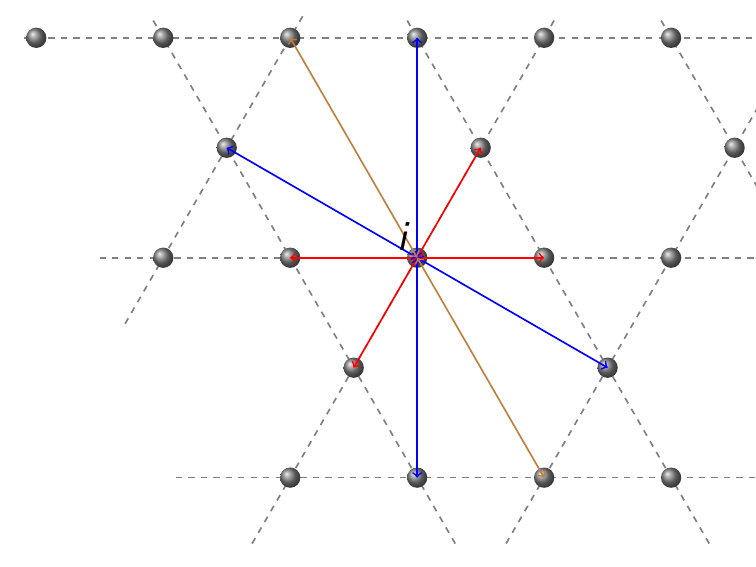}
\caption{Hoppings realized by the model Hamiltonian [Eq.~\eqref{eq:Ham}]: nearest-neighbor (red), next-nearest-neighbor (blue), and third-nearest-neighbor within hexagons (brown) from site $i$.}
\label{fig:hoppings}
\end{figure} 

\section{Results}
\subsection{Euler class}

A pair of isolated (gapped from the rest of the spectrum) bands $|u_{n}(\boldsymbol{k})\rangle, |u_{n+1}(\boldsymbol{k})\rangle$ can acquire a non-trivial Euler class $\chi$ when it is part of at least a three-band system that enjoys a reality condition assured by the presence of ${\cal C}_2\cal{T}$ [twofold rotations combined with time-reversal symmetry (TRS)], or ${\cal P} \cal{T}$ symmetry, involving parity and TRS. %This ensures that nodal degenacies between different bands with non-Abelian frame charge can induce non-triviality in the two-band subspace after braiding them in momentum space~\cite{bouhon2019nonabelian, BJY_nielsen, doi:10.1126/science.aau8740}.
The Euler class is then concretely obtained as~\cite{bouhon2019nonabelian, BJY_nielsen}
\begin{equation}\label{eq:Eulerpatch}
\chi= \dfrac{1}{2\pi}\int_{\text{BZ}}  \mathrm{Eu } ~\mathrm{d}k_1\wedge \mathrm{d}k_2 \in \mathbb{Z}, 
\end{equation}
where one integrates the Euler curvature $\mathrm{Eu} = \langle \partial_{k_1} u_n(\boldsymbol{k})\vert \partial_{k_2} u_{n+1}(\boldsymbol{k})\rangle - \langle \partial_{k_2} u_n(\boldsymbol{k})\vert \partial_{k_1} u_{n+1}(\boldsymbol{k})\rangle$ over the Brillouin zone (BZ). The pair of bands can either be degenerate (and flat) or feature a number of $2\chi$ nodal points that cannot be annihilated due to the topological nature~\cite{Clas7, bouhon2019nonabelian, BJY_nielsen}. Eq.~\eqref{eq:Eulerpatch} shows that the Euler class is the real analogue of the Chern number. Similarly, the isolated two-band subspace does not admit exponentially-localized Wannier functions in a ${\mathcal C}_2 \mathcal{T}$-symmetric gauge, but unlike the Chern case, the system does not feature protected chiral or helical edge states, allowing for a PEPS representation. Our PEPS construction will break the Wannier restriction to $\mathcal{C}_2\mathcal{T}$ symmetric gauge and three bands; more precisely, the latter, as it starts with virtual particles occupying six bands. The projection from virtual to physical fermions reduces the number of bands back to three, constituting a (non-invertible) map between the gauge-symmetry breaking input state and the gapped Euler state. Hence, a representation by a PEPS with a finite bond dimension is made possible by the PEPS construction breaking the gauge symmetry. 

\begin{figure}[b]
\includegraphics[width=0.35\textwidth]{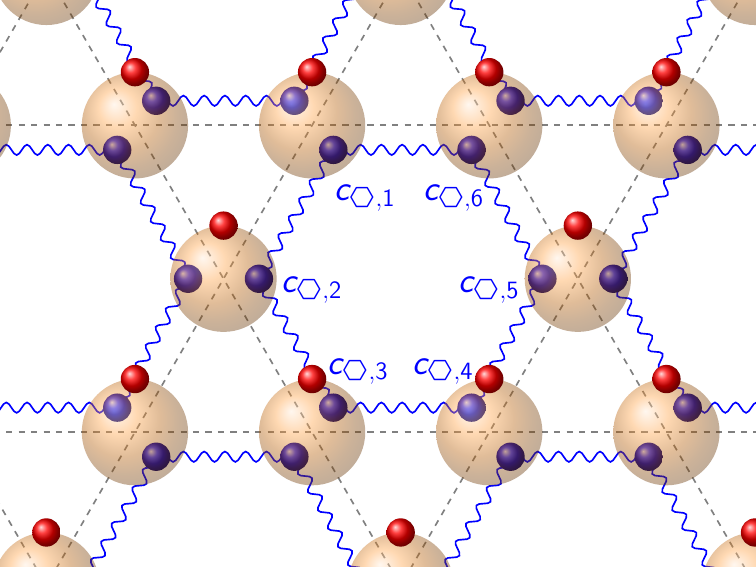}
\caption{Projected entangled simplex state. The blue wiggly lines denote the initial state of virtual fermions $c_{\hexagon,k}$ (blue balls) entangled across hexagons. The transparent red balls denote the projection onto the physical fermions (red balls).}
\label{fig:PESS}
\end{figure}

\subsection{The model}

To concretize the discussion, we consider spinless fermions hopping on the kagome lattice with nearest-neighbor hopping $t=-1$, next-nearest-neighbor hopping $t'=-1$ and third-nearest-neighbor hopping $t''=-1$ inside the hexagons (see Fig.~\ref{fig:hoppings}). For chemical potential $\mu$, the Hamiltonian thus reads
\begin{align}
    H = \sum_{\langle i,j\rangle} a_i^\dg a_j + \sum_{\langle \langle i,j \rangle \rangle} a_i^\dg a_j + \,\sum_{{\langle \langle \langle i,j \rangle \rangle \rangle_{\hexagon}}} a_i^\dg a_j - \mu \sum_{i=1}^N a_i^\dg a_i, \label{eq:Ham}
\end{align}
where $\langle i,j\rangle$, $\langle \langle i,j \rangle \rangle$ correspond to nearest- and next-nearest neighbor pairs of sites $i,j$ and $\langle \langle \langle i,j \rangle \rangle \rangle_{\hexagon}$ to third-nearest neighbor pairs of the same hexagons. $a_i^{\dg}$ ($a_i$) are the fermionic creation (annihilation) operators and $N$ the number of sites.  
The Hamiltonian has two degenerate flat bands at $E = -2-\mu$ and a dispersive band on top, separated by an energy gap $\Delta = 3$; for more details on the model, see Methods. The flat bands have Euler number $\chi = 1$, protected by $\mathcal{C}_2\mathcal{T}$ symmetry. At $\mu = -2$, both flat bands are at $E = 0$, and the ground states $|\psi_n\rangle, \ n = 0,1, \ldots, 2N/3$ are macroscopically degenerate, characterized by fillings $[0,1,2,\ldots, 2N/3]$. We now construct the ground state with the highest filling, which will become the unique ground state for $-2 < \mu < 1$. To that end, we note that the Hamiltonian for $\mu = -2$ can be rewritten as
\begin{align}\label{eq::Hhex}
H_p = \sum_{\hexagon} {\sum_{i,j \in \hexagon}} a_i^\dg a_j = 6 \sum_{\hexagon} a_{\hexagon}^\dg a_{\hexagon} = 6 \sum_{\hexagon} h_{\hexagon},
\end{align}
where $\hexagon$ denotes the hexagons of the kagome lattice, and we defined $a_{\hexagon} = \frac{1}{\sqrt{6}} {\sum_{i \in \hexagon}} a_i$ and $h_{\hexagon} = a_{\hexagon}^\dg a_{\hexagon}$. Hence, the ground states fulfill $a_{\hexagon} |\psi_n\rangle = 0$ for all hexagons $\hexagon$. The ground state with the highest occupation number is 
\begin{align}
|\psi_{2N/3} \rangle = \prod_{\hexagon} a_{\hexagon} | 1 \ldots 1\rangle,~\label{eq:constructPEPS}
\end{align}
where $|1 \ldots 1 \rangle$ is the fully occupied state. Due to $\{a_{\hexagon}, a_{\hexagon'}\}=0$, the ordering in Eq.~\eqref{eq:constructPEPS} is irrelevant. However, notably, for the other commutation relations, we have $\{a_{\hexagon}, a^\dg_{\hexagon'}\}=\delta_{\hexagon,\hexagon'}+\frac{1}{6} \delta_{\langle \hexagon,\hexagon' \rangle}$, where $\langle \hexagon,\hexagon' \rangle$ denotes corner-sharing, neighboring hexagonal plaquettes. This anticommutation algebra shows that, while the operators $a^\dagger_{\hexagon}$ effectively create fermions in a superposition of six atomic orbitals, the Hamiltonian Eq.~\eqref{eq::Hhex} is not adiabatically connected to a Hamiltonian of an atomic insulator, despite the functional similarity to such Hamiltonians: In particular, 
even if one deforms our quasiparticle operators $a_{\hexagon}$ (see following section), their non-trivial anticommutation properties are protected by $\mathcal{C}_2 \mathcal{T}$ symmetry. That is, the $a_{\hexagon}$ cannot be deformed into single-site operators (and similarly for the Hamiltonian) without breaking $\mathcal{C}_2 \mathcal{T}$ symmetry.
%that is the case due to the corner-sharing obstruction, the presence of which is also crucial for the entanglement of the system, as we demonstrate below. 

$|\psi_{2N/3}\rangle$ can be written as a projected entangled simplex state (PESS)~\cite{PESS} as follows: We start out with a virtual state of $2N$ spinless fermions -- two assigned to each physical fermion. The virtual fermions are in the state
%\begin{align}
    $|\omega_v\rangle = \frac{1}{\sqrt 6}\prod_{\hexagon} \sum_{i = 1}^6 c_{\hexagon,i} |1_v\rangle$, 
%\end{align}
where $|1_v \rangle$ denotes the fully occupied virtual state. $c_{\hexagon,i}$ corresponds to virtual fermion $i = 1, \ldots, 6$ within hexagon $\hexagon$ (as opposed to the physical particles $a_j$, there are now unique assignments within hexagons), cf. Fig.~\ref{fig:PESS}. The next step is to map each pair of virtual fermions around a site to one physical fermion. To that end, we use the operator $\hat M_j = a_j^\dg c_j' c_j + c_j' - c_j$. Here, $c_j'$ corresponds to the virtual fermion located on the left of the site $j$ and $c_j$ to the virtual fermion located on its right. We finally project on the vacuum of virtual particles, obtaining the overall state
\begin{align}
|\psi_\mr{PEPS} \rangle = \langle 0_v| \prod_{j=1}^N \hat M_j \prod_{\hexagon} \frac{1}{\sqrt 6} \sum_{i=1}^6 c_{\hexagon,i} |1_v 0_p\rangle,
\end{align}
where $|1_v 0_p\rangle$ corresponds to the vacuum of physical fermions and fully occupied virtual fermionic state. We already labeled the overall state as a ``PEPS'', since it can also be written as the more familiar projected entangled pair state, as we show further below. We note that it is a fermionic PEPS where tensors are replaced by fermionic operators~\cite{Kraus2010}. 
The map $\hat M_j = a_j^\dg c_j' c_j + c_j' - c_j$ has the following effect on the superposition of basis states in $\prod_{\hexagon} \frac{1}{\sqrt 6} \sum_{i=1}^6 c_{\hexagon,i} |1_v 0_p\rangle$: If site $j = \hexagon \cap \hexagon'$ is neither affected by $c_i$ with $i \in \hexagon$ nor $c_k$ with $k \in \hexagon'$ (i.e., there are two virtual fermions at site $j$), then a physical fermion is created via $a_j^\dg$. If site $j$ is affected by either $c_{i \in \hexagon}$ or $c_{k \in \hexagon'}$ (i.e., there is one virtual fermion at site $j$), then no physical fermion is created at site $j$, leaving it in the physical vacuum state. If site $j$ is affected by both $c_{i \in \hexagon}$ and $c_{k \in \hexagon'}$ (i.e., there is no virtual fermion at site $j$), the basis state is annihilated. The negative sign in $\hat M_j$ is necessitated by the fermionic anticommutation relations in the PEPS construction.

In order to demonstrate that $|\psi_{2N/3} \rangle \propto |\psi_\mr{PEPS}\rangle$, we first verify that $a_{\hexagon}|\psi_\mr{PEPS}\rangle = 0$ and later that the PEPS has filling $2N/3$. For the first claim, we notice that 
\begin{align}
&\langle 0_v| a_j (a_j^\dg c_j' c_j + c_j' - c_j) [\ldots] |1_v 0_p \rangle \notag \\
= &\langle 0_v | (a_j^\dg c_j' c_j + c_j' - c_j) c_j' [\ldots] |1_v 0_p\rangle  \notag \\
= &\langle 0_v| (a_j^\dg c_j' c_j + c_j' - c_j) c_j [\ldots] |1_v 0_p\rangle,
\end{align}
where $[\ldots]$ denotes a sum of products of operators that do not act on the physical fermion at site $j$. We thus have
\begin{align}
a_{\hexagon'} |\psi_\mr{PEPS}\rangle %= \pm \frac{1}{\sqrt 6} \langle 0_v | \sum_{i \in \hexagon'} a_i \prod_{j=1}^N \hat M_j \prod_{\hexagon} \sum_{k=1}^6 c_{\hexagon,k} |1_v 0_p\rangle \notag \\ 
&=\pm \frac{1}{6} \langle 0_v| \prod_j \left(a_j^\dg c_j' c_j + c_j' - c_j\right) \sum_{i=1}^6 c_{\hexagon',i} \times \notag \\
&\times \prod_{\hexagon} \sum_{k=1}^6 c_{\hexagon,k} |1_v 0_p\rangle = 0.    
\end{align}
Second, we see that the initial state $|1_v 0_p\rangle$ contains $2N$ virtual and no physical fermions. The operator $\prod_{\hexagon} \sum_{k=1}^6 c_{\hexagon,k}$ reduces that to $2N (1-1/6) = 5N/3$ fermions. Finally, each operator $\hat M_j$ creates one physical fermion less than it annihilates virtual ones, i.e., we are left with $2N/3$ physical fermions in $|\psi_\mr{PEPS}\rangle$. Hence, $|\psi_\mr{PEPS}\rangle \propto |\psi_{2N/3}\rangle$, as claimed. 

\begin{figure}[b]
\includegraphics[width=0.48\textwidth]{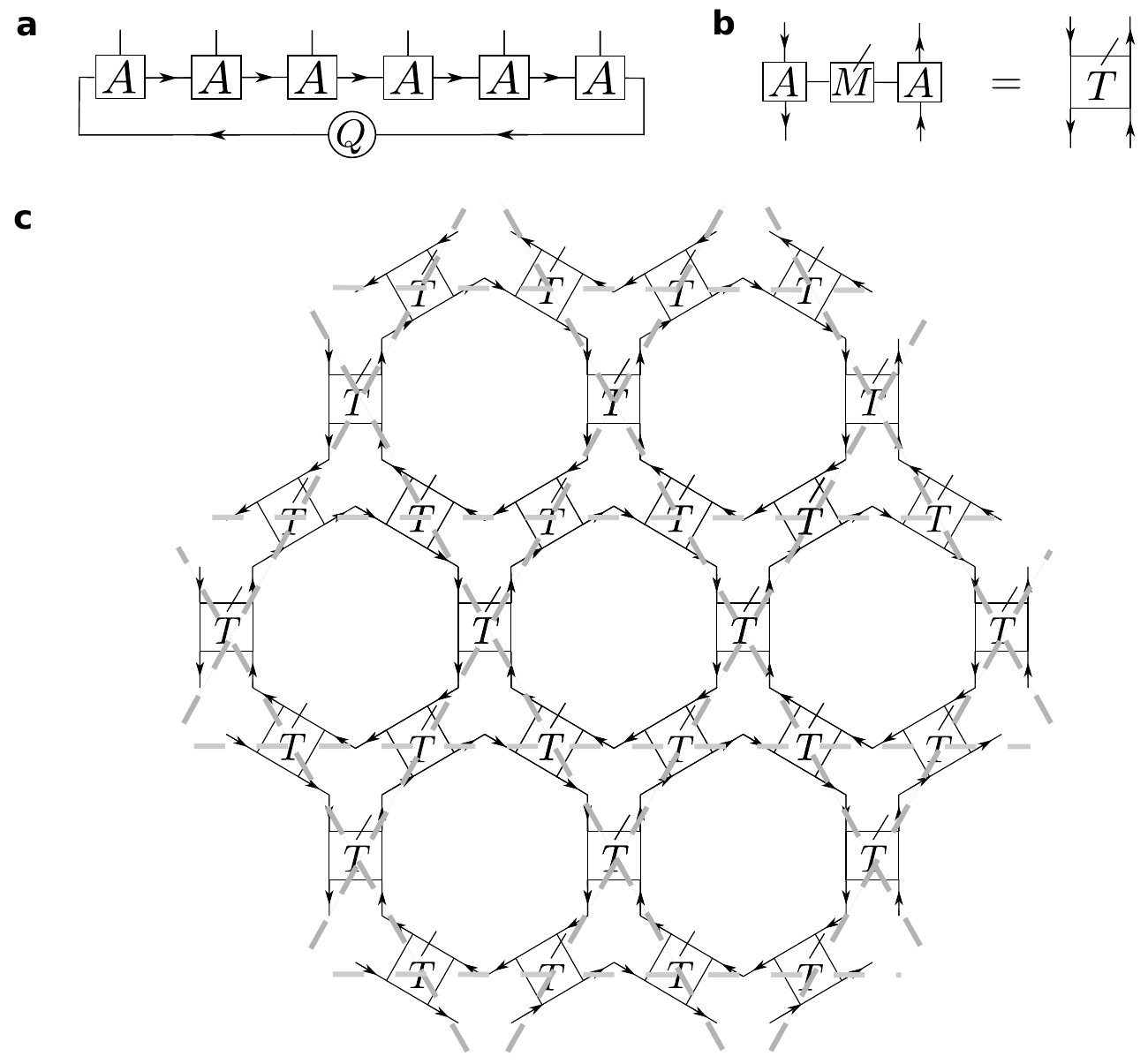}
\caption{PEPS construction. a: Matrix product state representation of the simplex states residing on the hexagons. $A$ can either be chosen to be of bond dimension $D = 2$, with $A^0_{12} = 1/\sqrt{6}, \,  A^1_{11} = -A^1_{22} = 1$, $Q_{21} = 1$ and all other elements of $A$ and $Q$ equal to zero, or $D = 6$ with $A^0_{61} = 1/\sqrt{6}, \, A^1_{l,l+1} = 1$ ($l = 1, \ldots, 5$), all other elements of $A$ equal to zero and $Q = \mathbb{1}$ (translationally invariant representation). Incoming arrows denote left and outgoing arrows right lower indices. b: By combining two $A$ tensors with the tensor $M$, we obtain the tensor $T$ constituting the PEPS. c: PEPS with one rank-5 tensor located on each site of the kagome lattice (gray dashed lines). }
\label{fig:PEPS}
\end{figure}

The PESS we have considered so far can be converted into a PEPS by realizing that the simplex states are of the form $|011111 \rangle + |101111\rangle + \ldots + |111110\rangle$, also known as a $W$-state~\cite{Duer2000}, which can be written as a non-translationally invariant matrix product state of bond dimension 2 or a translationally invariant one of bond dimension 6. $\hat M$ can be represented as a rank-3 tensor $M^i_{ab}$ with $M_{11}^1 = M_{10}^0 = -M_{01}^0 = 1$ and all other elements equal zero. The resulting PEPS tensor has rank 5 and bond dimension 2 or 6, respectively, see Fig.~\ref{fig:PEPS}. We note that a similar construction in terms of PESS defined on triangles can be used to describe the ground state of the Hamiltonian~\eqref{eq:Ham} for nearest-neighbor hopping only, an Euler system with one flat bottom band touched by two dispersive bands from above~\cite{Jiang_2021, JIANG2024}. % gapless \textcolor{orange}{in bottom bands? formualtion unclear} 

We now show that the PEPS is the unique ground state for $-2 < \mu < 1$: For chemical potential $\mu = -2$, the ground state subspace $S$ is the intersection of all null spaces of $h_{\hexagon}$, i.e., $S = \mr{span}\{\prod_{\hexagon} a_{\hexagon} a_{\hexagon}^\dg |n\rangle\}_n = \mr{span}\{\prod_{\hexagon} a_{\hexagon}|n\rangle\}_n$, where $\{|n\rangle\}_n$ is a complete basis and we used that $h_{\hexagon}^\perp = a_{\hexagon} a_{\hexagon}^\dg$ is a projector onto the orthogonal complement of $h_{\hexagon}$. Each $a_{\hexagon}$ eliminates one fermion from the basis state $|n\rangle$. Hence, the state with the highest occupation in $S$ is $\prod_{\hexagon} a_{\hexagon} |1 \ldots 1\rangle$, and it has filling $N - N/3 = 2N/3$. Any other state contained in $S$ has lower expectation value of the overall occupation. For $-2 < \mu < 1$, this highest occupation ground state becomes the unique ground state, as all other states contained in $S$ have lower overall occupation expectation value and thus get penalized.
%$|\psi_\mr{PEPS}\rangle$ is a (non-unique) frustration-free ground state of the parent Hamiltonian $H_p$ and the unique ground state of $H = H_p - (\mu + 2) \sum_{i=1}^N a_i^\dg a_i$ for $-2 < \mu < 1$.
This example shows that PEPS can be the unique ground states of local gapped Hamiltonians with non-trivial two-dimensional crystalline topological features, even in the non-interacting limit. This contrasts with the inability of free-fermionic PEPS to capture any higher-dimensional topological labels of the ten-fold classification unless the PEPS have algebraically decaying correlations~\cite{Read2017,Dubail2015,Wahl2013,Wahl2014}.

\subsection{Free-fermion generalizations} 

A straightforward generalization is obtained by modifying the simplex states to a linear combination, $|\tilde \omega_v\rangle = \prod_{\hexagon} \sum_{i=1}^6 \beta_i c_{\hexagon,i}|1_v\rangle$ with $\beta_i \in \mathbb{C}$, and keeping $\hat M_j$ the same. The new PEPS is annihilated by $\tilde a_{\hexagon} = \sum_{i \in \hexagon} \beta_i a_i$, where we set $\sum_{i=1}^6 |\beta_i|^2 = 1$. This corresponds to the new Hamiltonian
\begin{align}
    \tilde H = 6\sum_{\hexagon} \sum_{i,j \in \hexagon} \beta_i^* \beta_j a_i^\dg a_j - (\mu + 2) \sum_{i=1}^N a_i^\dg a_i.
\end{align}
One can check that $\mathcal{C}_2\mathcal{T}$ symmetry implies $\beta_{i+3}^* = \beta_i$ ($i = 1,2,3$) up to an irrelevant overall phase. 
For any choice of $\{\beta_i\}_{i=1}^6$, the corresponding PEPS is an Euler insulator, as the corresponding parent Hamiltonians are always gapped and continuously connected to the original one. Correspondingly, we have $\{\tilde a_{\hexagon}, \tilde a_{\hexagon'}^\dg\} = \delta_{\hexagon, \hexagon'}  + |\beta_{\hexagon \cap \hexagon'}|^2 \delta_{\langle \hexagon,\hexagon' \rangle}$, where there are always $i = \hexagon \cap \hexagon'$ for which the final term is non-vanishing. 
%Whether this state has a non-zero Euler number depends on the specific choice of the $\beta_i$. 

\begin{figure}[t]
\includegraphics[width=0.2\textwidth]{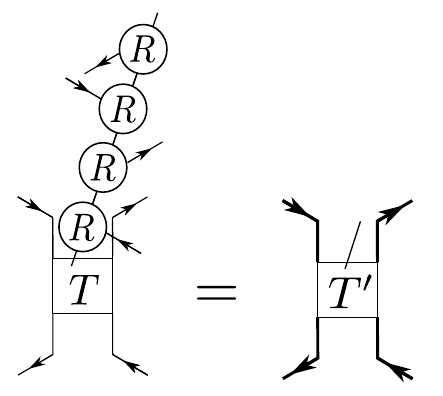}
\caption{Construction of the tensors $T'$ forming the building blocks of the interacting $|\psi_\mr{PEPS}'\rangle$. The $R$-tensors get absorbed into the $T$ tensor, increasing its bond dimension (indicated by thick directed lines). 
}
\label{fig:interactingPEPS}
\end{figure}

\subsection{Interacting generalizations} 

The modified simplex states give rise to non-interacting PEPS. We present two ways of generalizing these PEPS to the interacting regime. We will only briefly consider the first one, which employs a Gutzwiller projection and is likely to give rise to a new type of symmetry-enriched topological order~\cite{Chen2013}. The second example we consider applies symmetry-preserving shallow quantum circuits on the non-interacting PEPS. Therefore, it is guaranteed that the resulting interacting PEPSs are in the same topological phase as the original non-interacting one~\cite{Chen2010}.

To obtain the first type of interacting Euler state, we apply a Gutzwiller projection $P_G = \prod_i \left( a_{i \uparrow}^\dg a_{i\uparrow} a_{i \downarrow} a_{i \downarrow}^\dg +  a_{i \uparrow} a_{i\uparrow}^\dg a_{i \downarrow}^\dg a_{i \downarrow} \right) $ on two copies (denoted by $\uparrow$ and $\downarrow$) of the non-interacting PEPS, $|\psi_\mr{PEPS}^\uparrow\rangle$ and $\mathcal{PH}|\psi_\mr{PEPS}^\downarrow\rangle$, where $\mathcal{PH}$ induces a particle-hole transformation. This is necessary, as the Gutzwiller projector $P_G$ enforces one fermion per site, i.e., in order for it not to annihilate the state, we need overall filling fraction 1 in the two PEPS copies. The resulting state
\begin{align}
|\psi_\mr{PEPS}^{S = 1/2}\rangle = P_G |\psi^\uparrow_\mr{PEPS}\rangle \mathcal{PH} |\psi^\downarrow_\mr{PEPS} \rangle
\end{align}
is expected to be an Euler quantum spin liquid and to have fractional statistics, similarly to the Gutzwiller projection of two $p + ip$ superconducting states~\cite{Tu2013}. As the symmetries remain preserved under the Gutzwiller projection, $|\psi_\mr{PEPS}^{S = 1/2}\rangle$ is expected to be a symmetry-enriched topologically ordered state. This implies topological ground state degeneracy on the torus and anyonic excitations that survive breaking of the symmetry. If the protecting symmetry is not broken, additional features of non-interacting Euler insulators are expected, such as the cusp in the entanglement spectrum at $K = 0$ described below. However, studying the precise physical properties of this state goes beyond the scope of our work.

Second, we can also generalize the construction to interacting states by applying a shallow quantum circuit $U$ of diagonal unitaries on the non-interacting PEPS, which makes it easy to ensure that $\mathcal{C}_2\mathcal{T}$ symmetry is preserved. Hence, by definition, we remain in the same topological phase. Furthermore, the new state will also be a PEPS of low bond dimension. We consider the simplest case of nearest-neighbor gates. We view these as being applied on all hexagons in a translationally invariant fashion. Within each hexagon, we label $u_{j,j+1}$ as the unitary acting on sites $j$ and $j+1$ ($j = 7 \equiv 1$) inside a given hexagon, with sites enumerated as in Fig.~\ref{fig:PESS}. $\mathcal{C}_2\mathcal{T}$ symmetry is achieved if $u_{j,j+1} = u_{j+3,j+4}^*$ for all $j = 1,2,3$. The simplest continuously tuneable case is $u_{j,j+1} = \mathbb{1} - (1-e^{\pm i\alpha}) n_j n_{j+1}$ with particle number operators $n_j$, $\alpha \in [0,2\pi)$, and positive (negative) sign for $j = 1,2,3$ ($j = 4,5,6$). The new PEPS is given by $|\psi_\mr{PEPS}'\rangle = U |\psi_\mr{PEPS}\rangle$ and the Hamiltonian gets transformed as
\begin{align}
    H' = U H U^\dg = \sum_{\hexagon} \sum_{i,j \in \hexagon} {a_i'}^\dg a_j' - (\mu + 2) \sum_{i=1}^N a_i^\dg a_i,
\end{align}
where we defined $a_i' = U a_i U^\dg$ and used that $U$ commutes with $n_i = a_i^\dg a_i$. One can easily verify $a_i'^\dg = a_i^\dg \prod_{j}^{\langle i,j \rangle}[\mathbb{1} - (1-e^{i \sigma_{\langle i,j\rangle} \alpha}) n_j]$, where the product runs over all nearest neighbors of site $i$. $\sigma_{\langle i,j\rangle} = +1$ if $\langle i,j \rangle$ corresponds to one of the first three bonds in the hexagon that it lies in and $-1$ if it corresponds to one of the last three bonds. This gives rise to the overall Hamiltonian
\begin{align}
H' &= \sum_{\hexagon} \sum_{i,j \in \hexagon} a_i^\dg \prod_{k}^{\langle k,i \rangle}[\mathbb{1} - (1-e^{i \sigma_{\langle k,i\rangle} \alpha})n_k] \times \notag \\
&\times \prod_{l}^{\langle j,l \rangle}[\mathbb{1} - (1-e^{-i \sigma_{\langle j,l\rangle} \alpha}) n_l] \, a_j - (\mu + 2)\sum_{i=1}^N n_i.
\end{align}
$H'$ has the same spectrum as $H$ for fixed $\mu$, and $|\psi_\mr{PEPS}'\rangle$ is therefore its unique ground state for $-2 < \mu < 1$. The Hamiltonian is strictly local, acting on hexagons $\hexagon$ and adjacent triangles. Its four-body interactions have amplitude $\mathcal{O}(\alpha)$ and higher-body interactions are of higher order. 
This is the first example of an interacting Euler insulator with a local gapped Hamiltonian. $|\psi_\mr{PEPS}'\rangle$ can be constructed by writing the phase matrix of $u_{j,j+1} = \mathbb{1} - (1 - e^{\pm i \alpha})n_j n_{j+1}$ as $\sum_{q=1}^2 R^{ab}_q R^{cd}_q$ with $R^{ab}_1 = \delta_{ab}$ and $R^{ab}_2 = \sqrt{-1 + e^{\pm i \alpha}} \delta_{1a} \delta_{1b}$, $a,b \in \{0,1\}$. (As the underlying operators are even, they can be decomposed into tensor products.) Four nearest-neighbor unitaries act on each site, such that the tensors of $|\psi_\mr{PEPS}'\rangle$ can be constructed by contracting the physical leg of $T$ with four $R$-tensors, see Fig.~\ref{fig:interactingPEPS}. If the bond dimension of $T$ was chosen to be 2, the interacting PEPS has bond dimension $D = 4$.
We note that even though the protection of Euler topology under $\mathcal{C}_2\mathcal{T}$ symmetry in the presence of interactions is an open problem, in our case, we can guarantee that Euler topology is preserved,
as our interacting PEPS $|\psi_\mr{PEPS}'\rangle$ and interacting model Hamiltonian are both related via a symmetry-preserving
shallow quantum circuit to their non-interacting counterparts. A shallow quantum circuit does not change topological features, as these are global~\cite{Chen2010}. Hence, we are in the same symmetry-enriched topological phase~\cite{Chen2013} for all $\alpha \in [0,\pi]$. This can also be seen from the fact that the Hamiltonians $H'$ have the same energy spectrum (and gap) for all $\alpha$.

\subsection{Entanglement spectra}

\begin{figure}[t]
\includegraphics[width=0.49\textwidth]{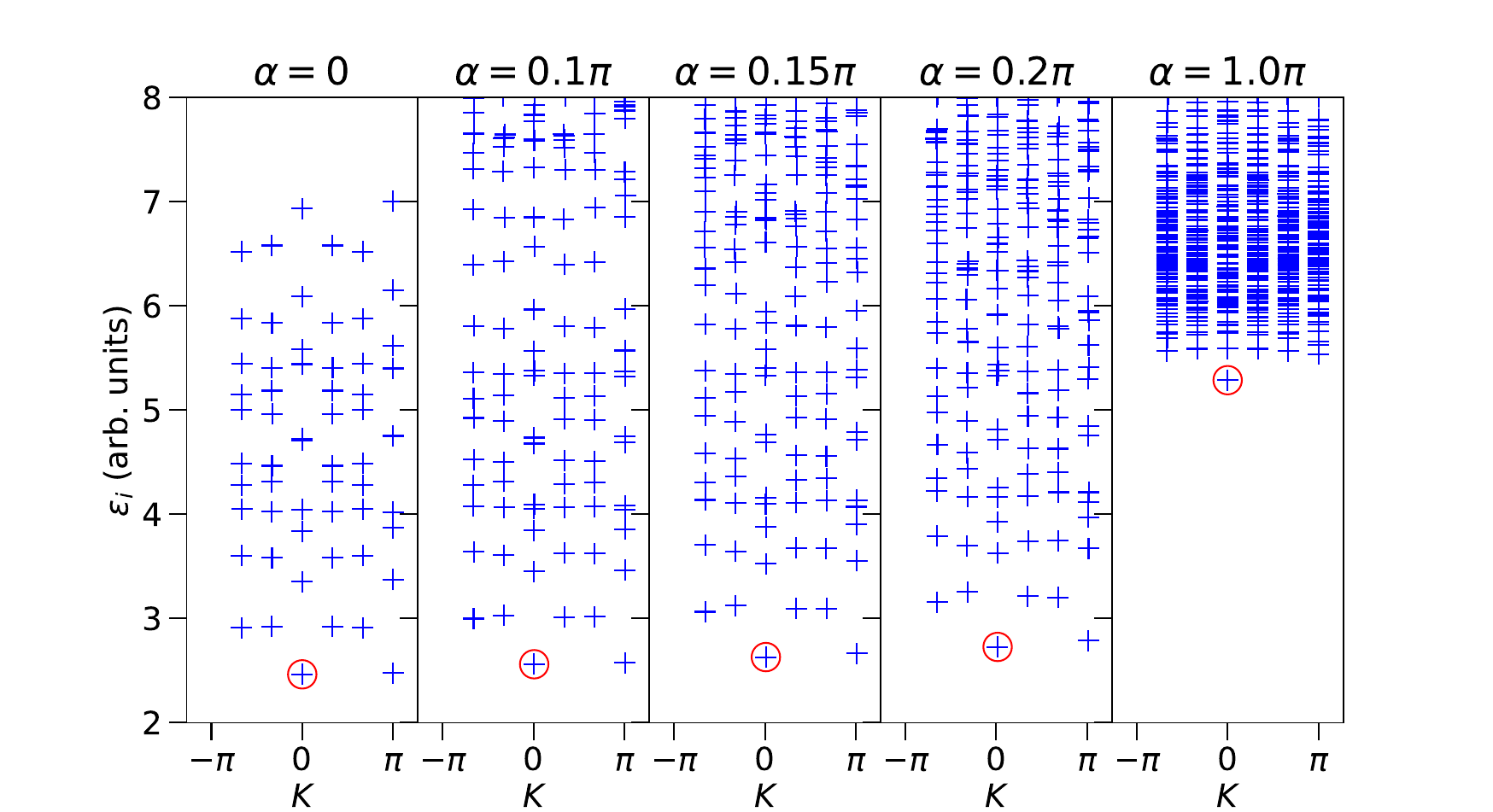}
\caption{Entanglement spectra as a function of the many-body momentum $K$ for different values of $\alpha$ for $L_y = 6$. For small values of $\alpha$, the low-lying spectrum strongly resembles the non-interacting one ($\alpha = 0$). In particular, a cusp at $K = 0$ (highlighted by a red circle)  is preserved as $\alpha$ is increased. 
Parallel to this, new entanglement energies appear at the top of the spectrum and eventually merge with its low-lying part.
}
\label{fig:ent_spectra}
\end{figure}

%\textcolor{orange}{We now discuss the entanglement features of the obtained PEPS. We begin with the non-interacting....mix in underneath for parts and new results of getting many body from non-interacting two particles..... We then move on to interacting. The spectrum may be traced as function of $\phi$ that shows that the levels spread but features of the central "star" survive.}
We numerically calculated the entanglement spectra of $|\psi_\mr{PEPS}'\rangle$ for an infinitely long torus with circumference $L_y$. That is, the torus is bipartitioned with $L_y$ unit cells located around the perimeter of the resulting cylinder. We obtained its entanglement spectrum by calculating the non-interacting $|\psi_\mr{PEPS}\rangle$ using TenPy~\cite{tenpy} and applying the quantum circuit $U$ on it to obtain the interacting $|\psi_\mr{PEPS}'\rangle$. The entanglement spectra for various values of $\alpha$ and $L_y = 6$ are shown in Fig.~\ref{fig:ent_spectra}. We observe that the low-lying part of the entanglement spectrum possesses a cusp at momentum $K = 0$, which remains intact as $\alpha $ is increased: Our non-interacting model corresponds to two degenerate Chern bands with opposite Chern number $C = \pm 1$. Due to continuous connection to the non-interacting limit, we expect that our $K = 0$ mode in the interacting case is not isolated, but connected to two isolated branches symmetric around the $K = 0$ axis  (gapless counter-propagating modes). Revealing these branches would require vastly larger system sizes, which are beyond the scope of our work. 
%This suggests that characteristic features of Euler insulators in the entanglement spectrum are preserved as interactions are turned on. 
%We note that in the more familiar case of Chern insulators, entanglement spectra are qualitatively preserved also only as long as interactions are weak. 
We further detail the entanglement features of the non-interacting case, including the stable cusp at $K = 0$, in the Methods.

% \textcolor{orange}{We need the two-particle and relation to entanglement. We can drop Majornana basis if needed Then we do the finite $\phi$-though we change that parameter...}

% Here, we comment on the entanglement spectra realized in the free fermion limit and in the interacting Hamiltonian, both of which are captured by the introduced construction. To describe the entanglement and further connect it to quantum geometry in the next section, we change the basis from complex fermions $a_i^\dg$, $a_i$ to Majorana basis $\gamma_{i,1/2}$. In the Majorana basis, we compute the correlation matrix $M_{ij}$, which with density matrix $\rho$ is given by
% %
% \begin{equation}
%     M_{ij} = \text{Tr} [i\rho \gamma_i\gamma_j],
% \end{equation}
% %
% in which we retrieve the correlations captured further by the quantum-geometric and quantum metrological bounds.

\section{Discussion} 

We leverage quantum geometric conditions (see Methods and SI) to define a class of exact PEPS with finite topological Euler invariant. The enigmatic nature of the Euler class allows to circumvent no-go conditions. Importantly, these models can be generalized to interacting variants and have definite entanglement signatures. As such, these PEPS set a benchmark for new 
pursuits. These potential pursuits involve studying exotic excitations and spin liquids realized from  Euler many-body PEPS ground states. In particular, on introducing interactions, novel kinds of fractionalizations should emerge from the interplay of the many-body entanglement as well as emergent quantum anomalous Hall states~\cite{bouhon2022multigap}. In addition, as all our states can be created by shallow quantum circuits from product states and have topological features, they are also particularly interesting for implementations on noisy intermediate-scale quantum devices and the development of new quantum error correction protocols. 
We will report on this in the near future.

\phantom{a}\\

\section{Methods}

\subsection{Momentum-space characterization of the model in the non-interacting limit}

We demonstrate how the model introduced in the main text, Eq.~(2), can be decomposed in momentum space. Furthermore, in the SI, we showcase the ideal non-Abelian quantum geometry realized in the topological Euler bands in the non-interacting limit. 

We~first Fourier transform the real-space creation and annihilation operators to the basis of Bloch orbitals:
$\tilde a^\dagger_{\alpha,\textbf{k}} = \frac{1}{\sqrt{N}} \sum_{l} a^{\dagger}_{\alpha,l} e^{- i \textbf{k} \cdot (\textbf{R}_l + \textbf{r}_{\alpha})} $, $\tilde a_{\alpha,\textbf{k}} = \frac{1}{\sqrt{N}} \sum_{l} a_{\alpha,l} e^{ i \textbf{k} \cdot (\textbf{R}_l + \textbf{r}_{\alpha})} $. Here, the operator $a^{(\dagger)}_{\alpha,l}$ annihilates (creates) a single particle in an atomic orbital $\alpha = A,B,C$ situated at position $\textbf{r}_{\alpha}$ with respect to the position vector of a unit cell center $\textbf{R}_l$, where $l = 1, 2, \ldots, N/3$ indexes unit cells. Due to the three sites per unit cell, we therefore have a three-band model. By inserting $a_{\alpha, l} = \frac{1}{2 \pi \sqrt N} \sum_{\mb k} \tilde a_{\alpha,\mb k} e^{-i \mb k \cdot (\mb R_l + \mb r_\alpha)}$ into Eq. (2) in the main text,  one obtains
\begin{equation}
    H = \sum_{\textbf{k};\alpha,\beta=A,B,C} \textit{H}_{\alpha \beta}(\textbf{k})~ \tilde a^{\dagger}_{\alpha,\textbf{k}} \tilde a_{\beta,\textbf{k}}.
\end{equation}
Here, the Bloch Hamiltonian for the considered system on the kagome lattice, manifestly expressed in a real gauge, reads~\cite{Jiang_2021, JIANG2024}
\begin{equation}
\textit{H}(\textbf{k}) = 
\begin{pmatrix}
        H_{AA}(\textbf{k}) & H_{AB}(\textbf{k}) & H_{AC}(\textbf{k})\\
        H_{AB}(\textbf{k}) & H_{BB}(\textbf{k}) & H_{BC}(\textbf{k})\\
        H_{AC}(\textbf{k}) & H_{BC}(\textbf{k}) & H_{CC}(\textbf{k}) \\
\end{pmatrix},
\\
\end{equation}
with the corresponding (real) matrix elements, on setting $t = t' = t'' = -1$ and $\mb k = (k_1,k_2)$,
\begin{align}
     & H_{AA}(\textbf{k}) = -\mu + 2 \cos{(k_1)}, \\
     & H_{AB}(\textbf{k}) =  2 \cos{(k_1/2+k_2/2)} + 2 \cos{(k_1/2-k_2/2)}, \\
     & H_{AC}(\textbf{k}) =  2 \cos{(k_2/2)} + 2 \cos{(k_1+k_2/2)}, \\
     & H_{BB}(\textbf{k}) = -\mu + 2 \cos{(k_2)}, \\
    & H_{BC}(\textbf{k}) =  2 \cos{(k_1/2)} + 2 \cos{(k_1/2+k_2)}, \\
    & H_{CC}(\textbf{k}) = -\mu + 2 \cos{(k_1+k_2)}.
\end{align}
%
%In the above, $\mu_{A}$, $\mu_{B}$, $\mu_{C}$ are the chemical potentials at the sublattices $A$, $B$, $C$, correspondingly, and we set $\mu = \mu_A = \mu_B = \mu_C$.

We recognize that the Bloch Hamiltonian can be further rewritten as
\begin{widetext}
\begin{equation}
\textit{H}(\textbf{k}) = 
\begin{pmatrix}
        -\mu - 2 + 4 \cos^2 (k_1/2)  & 4 \cos (k_1/2) \cos (k_2/2) & 4 \cos (k_1/2) \cos (k_1/2 + k_2/2)\\
        4 \cos (k_1/2) \cos (k_2/2) & -\mu - 2 + 4 \cos^2 (k_2/2)  & 4 \cos (k_2/2) \cos (k_1/2 + k_2/2)\\
        4 \cos (k_1/2) \cos (k_1/2 + k_2/2) & 4 \cos (k_2/2) \cos (k_1/2 + k_2/2) & -\mu - 2 + 4 \cos^2 (k_1 /2 + k_2/2) \\
\end{pmatrix},
\\
\end{equation}
\end{widetext}
or, more compactly,
\begin{equation}\label{eq::momH}
\textit{H}(\textbf{k}) = (-\mu - 2) \mathbb{1}_3 + 4 \textbf{n}(\textbf{k})~\otimes~\textbf{n}(\textbf{k})^{\text{T}},
\\
\end{equation}
with $\textbf{n}(\textbf{k}) = \Big( \cos (k_1/2), \cos (k_2/2), \cos (k_1/2 + k_2/2) \Big)^{\text{T}}$. Importantly, under such decomposition, the topology of the Euler bands in any three-band Hamiltonian satisfying a reality condition [$H(\textbf{k})=H^*(\textbf{k})$] can be captured by the normalized vector ${\hat{\textbf{n}}(\textbf{k}) = \textbf{n}(\textbf{k})/||\textbf{n}(\textbf{k})||}$. In particular, in the considered model, the vector $\hat{\textbf{n}}(\textbf{k})$ reads
\begin{align}
    \hat{\textbf{n}}(\textbf{k}) =
    &\frac{1}{\sqrt{\cos^2 (k_1/2) + \cos^2 (k_2/2) + \cos^2 (k_1/2 + k_2/2)}} \notag \\
    &\times 
    \begin{pmatrix}
        \cos (k_1/2) \\ 
        \cos (k_2/2) \\
        \cos (k_1/2 + k_2/2)
    \end{pmatrix},
\end{align}
and it fully determines the Euler curvature as
\begin{equation}
    \text{Eu} = \hat{\textbf{n}} \cdot (\partial_{k_2} \hat{\textbf{n}} \times \partial_{k_1} \hat{\textbf{n}}).
\end{equation}
The Euler curvature can be viewed as a skyrmion density in the momentum-space texture, with the skyrmion being spanned by $\hat{\textbf{n}}$ over the Brillouin zone (BZ) square/torus. In particular, the Euler invariant is given by~\cite{bouhon2019nonabelian, Jiang_2021}
\begin{equation}
    \chi = \frac{1}{2\pi} \int_{\text{BZ}} \dd^2 \textbf{k}~\text{Eu} = \frac{1}{2\pi} \int_{\text{BZ}} \dd^2 \textbf{k}~\hat{\textbf{n}} \cdot (\partial_{k_2} \hat{\textbf{n}} \times \partial_{k_1} \hat{\textbf{n}}) = 2 Q,
\end{equation}
and obtains $\chi = 1$ in the case of interest, which corresponds to the momentum-space meron (half-skyrmion) with the half-skyrmion number $Q = 1/2$~\cite{JIANG2024}. Additionally, the vector $\textbf{n}(\textbf{k})$ fully captures the band dispersion present in the model, as Eq.~\eqref{eq::momH} can be written as
\begin{equation}
\textit{H}(\textbf{k}) = (-\mu - 2) \mathbb{1}_3 + 4 ||\textbf{n}(\textbf{k})||^2 \hat{\textbf{n}}(\textbf{k})~\otimes~\hat{\textbf{n}}(\textbf{k})^{\text{T}},
\\
\end{equation}
explicitly determining the band dispersion in the third band as $E_3(\textbf{k}) = (-\mu - 2) + 4 ||\textbf{n}(\textbf{k})||^2$, contrary to the flat-band dispersion in the bottom Euler bands $E_1(\textbf{k}) = E_2(\textbf{k}) = (-\mu - 2)$. The band energies given by such dispersions manifestly have a gap across the entire Brillouin zone, as the norm of the vector $\textbf{n}(\textbf{k})$ is non-vanishing $||\textbf{n}(\textbf{k})|| > 0$ at every $\textbf{k}$-point. This follows from the fact that the components of the vector $\textbf{n}(\textbf{k})$, $\cos (k_1/2)$, $\cos (k_2/2)$, and $\cos (k_1/2 + k_2/2)$, are not independent, with at least one of those terms being necessarily non-vanishing at any $\textbf{k}$-point.

\begin{figure*}[t]
\includegraphics[width=1.0\textwidth]{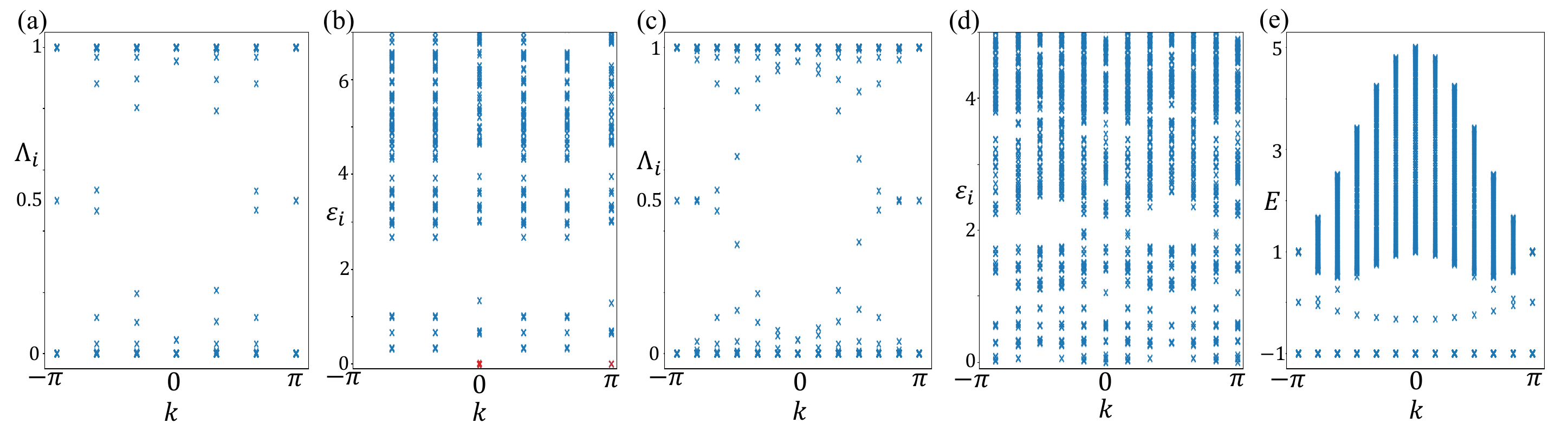}
\caption{ Entanglement, one-body correlation, and physical spectra: (a) One-body correlation spectrum $\Lambda_i$ on a thin torus with $L_y = 6$, (b) Many-body entanglement spectrum $\epsilon_i$ on an $L_y = 6$ torus. The red marker at $k=0$ and $\epsilon_i = 0$ is the ground state of the partition 'A'.
(c) and (d) are the one-body correlation spectrum $\Lambda_i$ and many-body entanglement spectrum $\epsilon_i$ for $L_y =12$, respectively, to clarify the variation along $k$.  (e) The physical spectrum $E$ on a cylinder of size $L_x = 120$ and $L_y = 12$ does not show an edge state with a spectral flow between the flat valence bands and dispersive conduction band.
}
\label{fig:ent_spectra_ni}
\end{figure*}

\subsection{Entanglement spectra of the non-interacting PEPS}

Here we present the entanglement spectrum of the non-interacting kagome Euler model of the main text. For this purpose, we start with the momentum space Hamiltonian defined on a thin torus, i.e., $L_x \gg L_y$. 
In the insulating state, the bottom two flat bands are occupied and the dispersive conduction band is empty. 
We first write down the projector on the occupied state
\begin{align}
    \hat{P}(\textbf{k}) = \sum_{i\in \text{occupied}} |\psi_i (\textbf{k})\rangle \langle \psi_i (\textbf{k}) |.
\end{align}
The projector by definition has its eigenvalues restricted to $0$ and $1$. For the calculations performed on a lattice, we define the real space positions as $ \textbf{r} = n_1 \textbf{a}_1 + n_2 \textbf{a}_2$, where $n_{1(2)} \in \mathbb{Z}$ and $\textbf{a}_{1(2)}$ are the lattice vectors.  
The corresponding reciprocal space momenta take the values as $ \textbf{k} = \frac{k_1}{2\pi} \textbf{b}_1 + \frac{k_2}{2\pi}\textbf{b}_2$, where $k_1, k_2 \in (-\pi,\, \pi]$ and $\textbf{b}_{1(2)}$ are the reciprocal lattice vectors. 
For the kagome model here, we have chosen, $\textbf{a}_1 = (\sqrt{3}/2,\, 1/2)$ and $\textbf{a}_2 = (0,\, 1)$ as the lattice vectors. The corresponding reciprocal lattice vectors are $\textbf{b}_1 = (4\pi/\sqrt{3},\, 0)$ and $\textbf{b}_2 = (-2\pi/\sqrt{3},\, 2\pi)$. 
From the projector, we obtain a one-body correlation operator 
\begin{align}
    G_{nm}(k_2) = \frac{1}{L_x}\sum_{k_1} \textbf{e}^{i 2\pi k_1 (n - m)} \hat{P}(k_1, k_2).
\end{align}
Since $G$ is also a projector, its eigenvalues are also restricted to $0$ and $1$. 
We partition the system into subsystems A and B, such that the entanglement spectrum between the two subsystems is given by the eigenvalues of the reduced density matrix $\rho_A$.  
The spectrum of the reduced density matrix $\rho_A$ can then be obtained from the spectrum of the reduced correlation matrix $G^A$ defined as~\cite{Peschel2003} 
\begin{align}
    G^A_{nm} (k_2) = G_{nm}(k_2);\quad n = 1, \ldots L_x, m = 1, \ldots, L_y.
\end{align}

In Fig.~\ref{fig:ent_spectra_ni} (a) and (c), we show the spectrum of the reduced one-body correlation matrix $G^A$. 
The plots are obtained for system sizes $L_x = 120$ and $L_y = 6 \, (12)$ for (a) and (c) respectively.
The eigenvalues $\Lambda_i(k_2)$ of $G^A$ are bounded to lie in $[0, \, 1]$, although, unlike the projector eigenvalues, they are not restricted to be $0$ and $1$. Indeed the in-gap eigenvalues are related to the topological Euler class of the model~\cite{Takahashi_2023}. 
However, unlike the well-known case of Chern insulators, these in-gap modes in the one-body correlation spectrum of the Euler topology are not related to the physical edge states due to non-trivial topology, which typically has a spectral flow between the bulk conduction and valence bands. 
In Fig.~\ref{fig:ent_spectra_ni} (e) we explicitly show the absence of such topological edge states with a spectral flow between flat valence bands at $-1$ and dispersive conduction band.  The physical energy spectrum is calculated for system size $L_x = 120$ and $L_y = 12$ with open boundary conditions along $L_x$ and periodic boundary conditions along $L_y$.

From the spectrum of $G^A$, we obtain the entanglement spectrum of the non-interacting model using the relation
\begin{align}\label{Eq:mbES}
    \varepsilon (k_2) =  &- \sum_{i \in \text{occupied} }\log [ \Lambda_i (k_2) ] \notag \\
    &-  \sum_{j \in \text{unoccupied}} \log [ 1 - \Lambda_j(k_2) ].
\end{align}
To calculate the full many-body entanglement spectrum as shown in the figure, we first obtain the ground state of subsection $A$ by occupying $2/3$ of the \textit{highest} eigenvalues $\Lambda_i$, which is commensurate with the $2/3$ filling of the whole system. 
Here, one should keep in mind that since the eigenvalues of the projector $\hat P$ of the occupied states lie at $1$ and unoccupied states at $0$, while constructing the ground state of $G^A$, one should start counting from $\Lambda_i \rightarrow 1$ as occupied, and going down in the eigenvalues $\Lambda_i$ corresponds to going up in the excitation spectrum. 
Once the ground state is identified, we obtain a many-body entanglement spectrum by creating excitations on this ground state. 
Since for the entanglement spectrum, we partition the system, while the filling fraction $2/3$ is a constraint for the whole system, the subsystem A has eigenstates that have particle numbers different to $2/3$ filling of the subsystem itself. 
Therefore to calculate the full many-body ground state, we first partition the subsystem A into different total particle number channels and from there create all possible particle-hole excitations. 
Then for each excited state configuration described by a fixed fermionic occupation number, we can obtain the entanglement spectrum using Eq.~\eqref{Eq:mbES}. 

The many-body entanglement spectrum of the non-interacting model is shown in Fig.~\ref{fig:ent_spectra_ni} (b) and (d) for the system size $L_x = 120$ and $L_y = 6 (12)$ respectively. 
We have taken the ground state entanglement energy to zero as a reference, which is shown by the red marker at $k = 0$ in Fig. ~\ref{fig:ent_spectra_ni} (b) and (d). 
Notice the presence of the zero entanglement energy state at $k=\pi$. This is obtained by considering a channel with one less (or more) particle in subsystem A than the exact $2/3$ filling. 

Comparing the many-body entanglement spectrum to the interacting case with $\alpha =0$ in the main text (left panel in Fig.~\ref{fig:ent_spectra}), we see a good agreement in their low energy features. In particular, in both cases, the ground state corresponds to $k=0$ with the lowest entanglement energy. As we move to a finite $k$, the entanglement energy increases and eventually comes back down to the ground state value at $k = \pi$ creating a cusp-like feature in the low-energy entanglement spectrum. 
This low energy behavior can be traced back to the in-gap modes (around $\Lambda = 0.5$) in the one-body correlation spectrum shown in Fig.~\ref{fig:ent_spectra_ni} (a) and (c). 
In the one-body correlation spectrum, for each mode near $0$, there are two modes near $1$, and therefore $2/3$ filling corresponds to occupying all modes in the upper half of the correlation spectrum. The low energy excitations are then created near $\Lambda_i = 0.5$ with a very low energy cost, which leads to many low energy modes in the entanglement spectrum.

%\textcolor{orange}{discuss features and cups and connect to o and $\pi$ counting and polarization modes of real boundary}

\section{Data Availability}

The datasets for the plots are available upon request.

\section{Code Availability}

The codes for our simulations are available upon request. They are based on the publicly available TeNPy library~\cite{tenpy}.

\section{Acknowledgements}

\begin{acknowledgments}
We acknowledge funding from a New Investigator Award, EPSRC grant EP/W00187X/1, an EPSRC ERC underwrite grant  EP/X025829/1, and a Royal Society exchange grant IES/R1/221060 as well as Trinity College, Cambridge (R.-J.S., G.C., and T.B.W.). This project was also supported by funding from the Rod Smallwood Studentship at Trinity College, Cambridge  (W.J.J.).
\end{acknowledgments}

\section{Author Contributions Statement}

R.-J.S. and T.B.W. initiated the project. T.B.W. constructed the PEPS and
performed the numerical simulations with input on the geometry and
speciﬁc models from R.-J.S, W.J.J., A.B and G.C. W.J.J. performed the
momentum-space characterisation of the model and quantum geometry with inputs from R.J.S., T.B.W., and A.B. G.C. numerically benchmarked the entanglement of the free fermion spectra. All authors
discussed the results. T.B.W. with input of R.-J.S. took the lead in the
writing of the manuscript, but the ﬁnal form of the manuscript beneﬁtted
from input from all authors.

\section{Competing Interests Statement}

The authors declare no competing interests.

\onecolumngrid
\setcounter{equation}{0}

\section{Supplementary Information for ``Exact projected entangled pair ground states with topological Euler invariant''}

\noindent In this Supplementary Information, we provide details on the quantum geometric properties of our non-interacting PEPS and connect the quantum Fischer information with the quantum Cramer-Rao bound for ideal Euler bands.

\section*{Supplementary Note 1: Quantum geometry in the free fermion limit}

In the following, we highlight the ideal quantum geometrical properties due to the flatness of the bottom bands. The flatness is crucial, as it allows for Hamiltonians which are sums of local projectors and therefore have (macroscopically degenerate) ground states at energy $E = 0$. 
The quantum metric~\cite{provost1980riemannian, bouhon2023quantum, tormaessay},
\begin{equation}
	g^{\chi}_{ij} = \text{Tr}_{\text{occ}} [(\partial_{k_i} \hat{P})(\partial_{k_j} \hat{P})],
\end{equation}
is defined as a trace over momentum-space projectors ${\hat{P} = \sum_{n=1,2} \ket{u_{n}(\textbf{k})} \bra{u_{n}(\textbf{k})}}$ of occupied Bloch states $\ket{u_{n}(\textbf{k})}$, with the momentum components ${k_i, k_j = k_1, k_2}$. The flat Euler bands saturate the quantum-geometric bounds, due to the non-zero Euler invariant $\chi$~\cite{bouhon2023quantum,kwon2024quantum}, between the quantum volume elements ($\sqrt{\text{det}~\textbf{g}^\chi}$) and the Euler curvature ($\sqrt{\text{det}~\textbf{g}^\chi} = |\text{Eu}|$) across the entire momentum space. Upon integrating, we thus retrieve a quantum volume which is a multiple of $2\pi$, 
\begin{equation}
	\text{Vol}~ \textbf{g}^\chi \equiv \oint \sqrt{\text{det}~\textbf{g}^\chi}~ \dd k_1 \wedge \dd k_2 = 2\pi |\chi|,
\end{equation}
showcasing the ideal non-Abelian quantum geometry~\cite{bouhon2023quantum, kwon2024quantum}. Upon introducing interactions, the many-body quantum metric in the space of twisted boundary conditions can reflect the topological nature of many-body Euler ground states as we also demonstrate in the non-interacting limit, see below. Using central relations of quantum metrology~\cite{PhysRevLett.72.3439, Liu_2020, SciPostMera2022}, the ideal condition physically manifests itself through the non-triviality of the quantum Fischer information (QFI). That is, we retrieve a metrological quantum Cramer-Rao (QCR) bound~\cite{PhysRevLett.72.3439} on the realizable model measurements, see the next Section, which could be directly executed in quantum simulators or synthetic three-level systems~\cite{Yu_2024}.

%Here, we elaborate on the quantum geometry of the model in non-interacting limit. 
We now concretize these points more formally. Consistently with the Pl\"ucker formalism for multi-band quantum geometry introduced in Ref.~\cite{bouhon2023quantum}, we can define the Fubini-Study metric (${\dd s^2 = 1 - |\braket{u_{1}(\textbf{k}) \wedge \ldots \wedge u_{n}(\textbf{k})|u_{1}(\textbf{k}+\dd \textbf{k}) \wedge \ldots \wedge u_{n}(\textbf{k}+\dd \textbf{k})}|^2}$) in the set of occupied Bloch bands $\{ \ket{u_{n}(\textbf{k})}\} $~\cite{provost1980riemannian},
\begin{equation}
	\dd s^2 = g^{\chi}_{ij}(\textbf{k}) \mathrm{d}k_i \mathrm{d}k_j
\end{equation}
where $g^{\chi}_{ij}$ is the quantum metric in the Euler flat bands, and the Einstein summation convention was assumed. With both of the Euler bands $n=1,2$ occupied (`occ'), we can correspondingly write the metric as
\begin{equation}\label{eq:metric_def}
	g^{\chi}_{ij}(\textbf{k}) = \sum^\text{occ}_{n} \frac{1}{2} \Big[ \bra{\partial_{k_i} u_{n}(\textbf{k})} \hat{Q} \ket{\partial_{k_j} u_{n}(\textbf{k})} + \text{c.c.} \Big] = \braket{\partial_{k_i} u_{1}(\textbf{k})|u_{3}(\textbf{k})}  \braket{u_{3}(\textbf{k})|\partial_{k_j} u_{1}(\textbf{k})} + \braket{\partial_{k_i} u_{2}(\textbf{k})|u_{3}(\textbf{k})}  \braket{u_{3}(\textbf{k})|\partial_{k_j} u_{2}(\textbf{k})},
\end{equation}
where $\hat{Q} = \sum^\text{unocc}_{m} \ket{u_{m}(\textbf{k})}  \bra{u_{m}(\textbf{k})} = \ket{u_{3}(\textbf{k})}  \bra{u_{3}(\textbf{k})}$ is the projector onto unoccupied (`unocc') band(s); here, $m=3$. In the second equality, we used the fact that the eigenvectors representing the Euler bands are chosen real, as here, the Bloch Hamiltonian $H(\textbf{k}) = \sum^3_{i=1} E_{i}(\textbf{k}) \ket{u_{i}(\textbf{k})} \bra{u_{i}(\textbf{k})}$ is a real symmetric matrix. The quantum metric is manifestly real and symmetric, by definition Supplementary Equation~\eqref{eq:metric_def}.

Alternatively, we can rewrite the metric in terms of the projector onto the unoccupied band as,
\begin{align}
	\begin{split}
		g^\chi_{ij} = \text{Tr}_{\text{occ}} [(\partial_{k_i} \hat{Q})(\partial_{k_j} \hat{Q})] = \text{Tr}_{\text{occ}} (\ket{\partial_{k_i} u_{3}(\textbf{k})} \bra{\partial_{k_j} u_{3}(\textbf{k})} + \ket{u_{3}(\textbf{k})}\braket{u_{3}(\textbf{k})| \partial_{k_i} 
			u_{3}(\textbf{k})}\bra{\partial_{k_j} u_{3}(\textbf{k})} \\ + \ket{u_{3}(\textbf{k})}\braket{\partial_{k_i} u_{3}(\textbf{k})|u_{3}(\textbf{k})}\bra{\partial_{k_j} u_{3}(\textbf{k})} + \ket{u_{3}(\textbf{k})}\braket{\partial_{k_i} u_{3}(\textbf{k})|\partial_{k_j} u_{3}(\textbf{k})}\bra{u_{3}(\textbf{k})}) \\= \braket{\partial_{k_i} u_{3}(\textbf{k}) | \partial_{k_j} u_{3}(\textbf{k})} - \braket{\partial_{k_i} u_{3}(\textbf{k})|
			u_{3}(\textbf{k})}\braket{u_{3,\textbf{k}
			}|\partial_{k_j} u_{3,\textbf{k}
		}} = \braket{\partial_{k_i} u_{3}(\textbf{k}) | \partial_{k_j} u_{3}(\textbf{k})} 
	\end{split}
\end{align}
where the last equality follows from the reality condition. We now us the fact that the third Bloch band defines a normalized vector field: $\hat{\textbf{n}}(\textbf{k}) \hat{=} \ket{u_{3}(\textbf{k})}$, as follows from the spectral decomposition of the Hamiltonian.

In terms of the momentum-space vector $\hat{\textbf{n}}$, the quantum metric in a three-band Euler Hamiltonian reads~\cite{jankowski2023optical}
\begin{equation}
	g^{\chi}_{ij} =  (\partial_{k_i} \hat{\textbf{n}}) \cdot (\partial_{k_j} \hat{\textbf{n}}),
\end{equation}
which obtains an inequality~\cite{kwon2024quantum},
\begin{equation}
	\sqrt{\text{det}~\textbf{g}^\chi} \geq |\text{Eu}|.
\end{equation}
Additionally, from inequality between arithmetic and geometric means, we directly obtain,
\begin{equation}
	\text{Tr}~ \textbf{g}^\chi \equiv g^{\chi}_{11} + g^{\chi}_{22} \geq 2 \sqrt{g^{\chi}_{11} g^{\chi}_{22}} \geq 2 \sqrt{g^{\chi}_{11} g^{\chi}_{22} - (g^{\chi}_{12})^2} \equiv 2 \sqrt{\text{det}~\textbf{g}^\chi} \geq 2|\text{Eu}|,
\end{equation}
where we used the symmetry of the (real) metric tensor $g^{\chi}_{12} = g^{\chi}_{21}$.

In the considered model, the metric elements read:
\begin{equation}
	g^{\chi}_{11} = \frac{8 - 3 \cos k_1 - 3\cos (k_1 + k_2) - \cos (k_1 - k_2) - \cos (k_1 + 2 k_2)}{8(3 + \cos k_1 + \cos k_2 + \cos(k_1 + k_2))^2},
\end{equation}
\begin{equation}
	g^{\chi}_{22} = \frac{8 - 3 \cos k_2 - 3\cos (k_1 + k_2) - \cos (k_1 - k_2) - \cos (2 k_1 + k_2)}{8(3 + \cos k_1 + \cos k_2 + \cos(k_1 + k_2))^2},
\end{equation}
\begin{equation}
	g^{\chi}_{12} = g^{\chi}_{21} = \frac{2 - 2 \cos k_1 \cos k_2 + \sin k_1 \sin k_2 }{4(3 + \cos k_1 + \cos k_2 + \cos(k_1 + k_2))^2},
\end{equation}
which directly obtains the quantum volume~\cite{kwon2024quantum},
\begin{equation}
	\sqrt{\text{det}~\textbf{g}^\chi (\textbf{k})}  = \frac{-3 + \cos k_1 + \cos k_2 + \cos (k_1 + k_2)}{4\sqrt{2} (3 + \cos k_1 + \cos k_2 + \cos(k_1 + k_2))^{3/2}},
\end{equation}
as well as
\begin{equation}
	\text{Tr}~ \textbf{g}^\chi (\textbf{k}) = \frac{16 - 3 \cos k_1 - 3 \cos k_2 - 6 \cos (k_1 + k_2) - 2 \cos (k_1 - k_2) - \cos (2 k_1 + k_2) - \cos (k_1 + 2 k_2)}{8(3 + \cos k_1 + \cos k_2 + \cos(k_1 + k_2))^2}.
\end{equation}
On the contrary, the Euler curvature in the model is given by the following expression
\begin{equation}
	\text{Eu}(\textbf{k}) = \frac{-3 + \cos k_1 + \cos k_2 + \cos (k_1 + k_2)}{4\sqrt{2} (3 + \cos k_1 + \cos k_2 + \cos(k_1 + k_2))^{3/2}}.
\end{equation}
We note that, analytically, an inequality $\text{Tr}~ \textbf{g}^\chi (\textbf{k}) \geq 2|\text{Eu}(\textbf{k})|$ holds, on substituting the individual quantum metric matrix elements to the bound between the determinant and trace. The equality of the determinant (quantum volume) and the Euler curvature follows trivially by inspection, as the analytical expressions for both quantities are identical across the entire momentum space.

Moreover, beyond the single-particle context, we can consider many-body quantum metric $g_{ij}(\mathrm{\theta})$ defined in terms of the twist angles $\mathrm{\theta} = (\theta_1, \theta_2)$ and the twisted boundary conditions~\cite{PhysRevB.31.3372},
\begin{equation}
	\psi(\{x_i + L_1\}, \{y_i\}) \equiv \braket{\{x_i + L_1\}, \{y_i\} | \psi} = e^{i \theta_1} \psi(\{x_i\}, \{y_i\}),
\end{equation}
\begin{equation}
	\psi(\{x_i\}, \{y_i + L_2\}) \equiv \braket{\{x_i\}, \{y_i  + L_2\} | \psi} = e^{i \theta_2} \psi(\{x_i\},\{ y_i\}),
\end{equation}
where $i = 1, 2, \ldots, N_{\text{tot}}$ are particle labels, $x_i$ and $y_i$ the sets of coordinates of fermions $i$, and $L_1, L_2$ denote the cell lengthscales, on which the twisted periodic boundary conditions were imposed on the many-body state.

In terms of the twist angles, the many-body quantum metric reads
\begin{equation}
	g_{ij}(\mathrm{\theta}) = \mathfrak{Re}~ \bra{\partial_{\theta_i} \psi(\mathrm{\theta})} (1 - \hat{P}_\theta) \ket{\partial_{\theta_j} \psi(\mathrm{\theta})},
\end{equation}
with the projector onto the many-body ground state $\hat{P}_\theta = \ket{\psi(\mathrm{\theta})} \bra{\psi(\mathrm{\theta})}$.
At the zero twist angle $\mathrm{\theta} = (0, 0) \equiv \mathbf{0}$, in the free-fermion limit, we retrieve a many-body bound,
\begin{equation}
	g_{ij}(\mathbf{0}) = \frac{1}{L_1 L_2} \sum_{\textbf{k}} \text{Tr}~ \textbf{g}^\chi (\textbf{k}) \geq \frac{4\pi A}{L_1 L_2} |\chi|,
\end{equation}
where $A$ is the area of the unit cell of the system. However, we note that unlike in the case of the determinant bound providing an ideal condition on Euler bands, this many-body bound does not saturate in the considered models, as it reduces to the trace bound in the free particle limit. In other words, here, the strong inequalities rather than equalities hold within the proposed models.

\section*{Supplementary Note 2: Quantum Fischer information and quantum Cramer-Rao bound of the ideal Euler bands}
We further comment on the structures present in the models and their relation to quantum Fisher information (QFI) and the quantum Cramer-Rao (QCR) bound~\cite{PhysRevLett.72.3439, Liu_2020}. Namely, we derive a non-Abelian QCR bound, which is induced by Euler topology and the Euler bands satisfying an ideal condition.
The QCR bounds are of central relevance for quantum metrology~\cite{Liu_2020}.

To define the QFI in the context of this work, we consider a two-parameter family of single-particle states $\ket{\psi_{n}(\textbf{k})}$, parametrized by $\textbf{k} \equiv (k_1, k_2) \in T^2$, where $T^2$ denotes a two-torus. Consistently with the models introduced in the main text, at every point $\textbf{k}$ of the parameter space, we consider a three-state system. We take a spectral decomposition of the density matrix of the single-particle states at given $\textbf{k}$-point, $\rho \equiv \sum_n \lambda_n \ket{\psi_{n}(\textbf{k})}\bra{\psi_{n}(\textbf{k})}$. The QFI matrix for single-particle operators $\hat{r}_1$, $\hat{r}_2$ (conjugate to $k_1, k_2$), then reads~\cite{Liu_2020},
\begin{equation}
	F_{ij}[\rho] \equiv \sum_{m,n: \lambda_m + \lambda_n \geq 0} 2\frac{(\lambda_m - \lambda_n)^2}{\lambda_m + \lambda_n} \bra{\psi_{m}(\textbf{k})} \hat{r}_i \ket{\psi_{n}(\textbf{k})}  \bra{\psi_{n}(\textbf{k})} \hat{r}_j \ket{\psi_{m}(\textbf{k})}.
\end{equation}
In translationally symmetric contexts, if the parameters $\textbf{k}$ were to be identified with momenta, then $\hat{r}_1$, $\hat{r}_2$ represent position operator components defined along the lattice vectors. We moreover recognize: $\hat{r}_i \sim i\partial_{k_i}$, i.e. ${-i \partial_{k_i} \rho = [\rho, \hat{r}_i]}$, and hence, $-i \bra{\psi_{m} (\textbf{k})} \partial_{k_i} \rho  \ket{\psi_{n} (\textbf{k})} = \bra{\psi_{m} (\textbf{k})} [\rho, \hat{r}_i] \ket{\psi_{n}(\textbf{k})} = (\lambda_m - \lambda_n) \bra{\psi_{m}(\textbf{k})} \hat{r}_i \ket{\psi_{n} (\textbf{k})}$. Therefore, for pure states, where in the context of this work we can consider single particle in the third band within the corresponding three-state problem, i.e. ${\rho = \ket{\psi_3(\textbf{k})}\bra{\psi_3(\textbf{k})} = \ket{u_3(\textbf{k})}\bra{u_3(\textbf{k})}}$; the QFI matrix reduces to the quantum metric (see the main text),
\begin{equation}
	F_{ij}[\rho] = 4 g^{\chi}_{ij}(\textbf{k}).
\end{equation}
The QCR bound~\cite{PhysRevLett.72.3439, Liu_2020} for the two-parameter measurements can be captured by the covariance matrix $\Sigma$ with~\cite{Liu_2020, Yu_2024},
\begin{equation}
	\Sigma (\hat{\textbf{k}}) \geq \frac{1}{M} F^{-1} [\rho],
\end{equation}
where $M$ is the number of the repetitions of measurements~\cite{Liu_2020, SciPostMera2022, Yu_2024}. The covariance matrix for an unbiased estimator $\hat{\textbf{k}}$ for the two-parameter family $\textbf{k} = (k_1, k_2)$ under a set of positive operator-valued measurements (POVM), ${\Pi_p }$, such that $\sum^{N_p}_p \Pi_p = 1$, $\Pi_p \Pi_{p'} = \Pi_p \delta_{p p'}$, with $N_p \geq 3$; is defined as~\cite{Liu_2020, SciPostMera2022},
\begin{equation}
	\Sigma_{ij} (\hat{\textbf{k}}) = \langle \delta k_i \delta k_j \rangle \equiv \sum_{p} k_i k_j \text{Tr}[\rho \Pi_p] - k_i k_j,
\end{equation}
where $\langle \ldots \rangle \equiv \text{Tr}[ \rho(\ldots) ]$. For ideal bands, as in the introduced model, the Euler curvature determines the quantum-metrological bound at every point of the parameter space as,
\begin{equation}
	\sqrt{\text{det}~ \Sigma(\hat{\textbf{k}})} \geq  \frac{1}{M \sqrt{\text{det}~\textbf{g}^{\chi}(\textbf{k})}} = \frac{1}{M | \text{Eu}(\textbf{k})|},
\end{equation}
where the first inequality follows from the derivation of Ref.~\cite{SciPostMera2022}, and the second equality is realized in the models introduced in our work. 

%Notably, such bound involving non-Abelian Berry curvature, is a generalization of a topological bound due to Abelian Berry curvature, i.e. diagonal parts of the non-Abelian Berry curvature, which was experimentally measured~\cite{Yu_2024}. However, in the Hamiltonians hosting the non-Abelian Euler invariant, as considered in this work, the sum of the diagonal elements of non-Abelian Berry curvature over occupied bands vanishes identically, unlike in the ideal Chern bands. Hence, the quantum-metrological bound derived here is manifestly distinct. 

Beyond the demonstrated quantum-metrological manifestations, the realized ideal condition for the Euler bands opens avenues for exotic fractionalization of the excitations in the topological bands, and offers a platform for exploring further deeper connections to the many-body quantum metric under twisted boundary conditions.


\begin{thebibliography}{63}%
	\makeatletter
	\providecommand \@ifxundefined [1]{%
		\@ifx{#1\undefined}
	}%
	\providecommand \@ifnum [1]{%
		\ifnum #1\expandafter \@firstoftwo
		\else \expandafter \@secondoftwo
		\fi
	}%
	\providecommand \@ifx [1]{%
		\ifx #1\expandafter \@firstoftwo
		\else \expandafter \@secondoftwo
		\fi
	}%
	\providecommand \natexlab [1]{#1}%
	\providecommand \enquote  [1]{``#1''}%
	\providecommand \bibnamefont  [1]{#1}%
	\providecommand \bibfnamefont [1]{#1}%
	\providecommand \citenamefont [1]{#1}%
	\providecommand \href@noop [0]{\@secondoftwo}%
	\providecommand \href [0]{\begingroup \@sanitize@url \@href}%
	\providecommand \@href[1]{\@@startlink{#1}\@@href}%
	\providecommand \@@href[1]{\endgroup#1\@@endlink}%
	\providecommand \@sanitize@url [0]{\catcode `\\12\catcode `\$12\catcode
		`\&12\catcode `\#12\catcode `\^12\catcode `\_12\catcode `\%12\relax}%
	\providecommand \@@startlink[1]{}%
	\providecommand \@@endlink[0]{}%
	\providecommand \url  [0]{\begingroup\@sanitize@url \@url }%
	\providecommand \@url [1]{\endgroup\@href {#1}{\urlprefix }}%
	\providecommand \urlprefix  [0]{URL }%
	\providecommand \Eprint [0]{\href }%
	\providecommand \doibase [0]{https://doi.org/}%
	\providecommand \selectlanguage [0]{\@gobble}%
	\providecommand \bibinfo  [0]{\@secondoftwo}%
	\providecommand \bibfield  [0]{\@secondoftwo}%
	\providecommand \translation [1]{[#1]}%
	\providecommand \BibitemOpen [0]{}%
	\providecommand \bibitemStop [0]{}%
	\providecommand \bibitemNoStop [0]{.\EOS\space}%
	\providecommand \EOS [0]{\spacefactor3000\relax}%
	\providecommand \BibitemShut  [1]{\csname bibitem#1\endcsname}%
	\let\auto@bib@innerbib\@empty
	%</preamble>
	\bibitem [{\citenamefont {Verstraete}\ and\ \citenamefont
		{Cirac}(2006)}]{Verstraete2006}%
	\BibitemOpen
	\bibfield  {author} {\bibinfo {author} {\bibfnamefont {F.}~\bibnamefont
			{Verstraete}}\ and\ \bibinfo {author} {\bibfnamefont {J.~I.}\ \bibnamefont
			{Cirac}},\ }\bibfield  {title} {\bibinfo {title} {Matrix product states
			represent ground states faithfully},\ }\href
	{https://doi.org/10.1103/physrevb.73.094423} {\bibfield  {journal} {\bibinfo
			{journal} {Phys. Rev. B}\ }\textbf {\bibinfo {volume} {73}},\ \bibinfo
		{pages} {094423} (\bibinfo {year} {2006})}\BibitemShut {NoStop}%
	\bibitem [{\citenamefont {Huang}(2014)}]{Huang2014}%
	\BibitemOpen
	\bibfield  {author} {\bibinfo {author} {\bibfnamefont {Y.}~\bibnamefont
			{Huang}},\ }\bibfield  {title} {\bibinfo {title} {Area law in one dimension:
			Degenerate ground states and {R}enyi entanglement entropy},\ }\href
	{https://arxiv.org/abs/1403.0327} {\bibfield  {journal} {\bibinfo  {journal}
			{Preprint at https://arxiv.org/abs/1403.0327}\ } (\bibinfo {year}
		{2014})}\BibitemShut {NoStop}%
	\bibitem [{\citenamefont {Molnar}\ \emph {et~al.}(2015)\citenamefont {Molnar},
		\citenamefont {Schuch}, \citenamefont {Verstraete},\ and\ \citenamefont
		{Cirac}}]{Molnar2015}%
	\BibitemOpen
	\bibfield  {author} {\bibinfo {author} {\bibfnamefont {A.}~\bibnamefont
			{Molnar}}, \bibinfo {author} {\bibfnamefont {N.}~\bibnamefont {Schuch}},
		\bibinfo {author} {\bibfnamefont {F.}~\bibnamefont {Verstraete}},\ and\
		\bibinfo {author} {\bibfnamefont {J.~I.}\ \bibnamefont {Cirac}},\ }\bibfield
	{title} {\bibinfo {title} {{Approximating Gibbs states of local Hamiltonians
				efficiently with projected entangled pair states}},\ }\href
	{https://doi.org/10.1103/physrevb.91.045138} {\bibfield  {journal} {\bibinfo
			{journal} {Phys. Rev. B}\ }\textbf {\bibinfo {volume} {91}},\ \bibinfo
		{pages} {045138} (\bibinfo {year} {2015})}\BibitemShut {NoStop}%
	\bibitem [{\citenamefont {Dalzell}\ and\ \citenamefont
		{Brandão}(2019)}]{Dalzell2019}%
	\BibitemOpen
	\bibfield  {author} {\bibinfo {author} {\bibfnamefont {A.~M.}\ \bibnamefont
			{Dalzell}}\ and\ \bibinfo {author} {\bibfnamefont {F.~G. S.~L.}\ \bibnamefont
			{Brandão}},\ }\bibfield  {title} {\bibinfo {title} {{Locally accurate MPS
				approximations for ground states of one-dimensional gapped local
				Hamiltonians}},\ }\href {https://doi.org/10.22331/q-2019-09-23-187}
	{\bibfield  {journal} {\bibinfo  {journal} {Quantum}\ }\textbf {\bibinfo
			{volume} {3}},\ \bibinfo {pages} {187} (\bibinfo {year} {2019})}\BibitemShut
	{NoStop}%
	\bibitem [{\citenamefont {Huang}(2019)}]{Huang2019}%
	\BibitemOpen
	\bibfield  {author} {\bibinfo {author} {\bibfnamefont {Y.}~\bibnamefont
			{Huang}},\ }\bibfield  {title} {\bibinfo {title} {Approximating local
			properties by tensor network states with constant bond dimension},\ }\href
	{https://arxiv.org/abs/1903.10048} {\bibfield  {journal} {\bibinfo  {journal}
			{Preprint at https://arxiv.org/abs/1903.10048}\ } (\bibinfo {year}
		{2019})}\BibitemShut {NoStop}%
	\bibitem [{\citenamefont {White}\ and\ \citenamefont
		{Scalapino}(1998)}]{White1998}%
	\BibitemOpen
	\bibfield  {author} {\bibinfo {author} {\bibfnamefont {S.~R.}\ \bibnamefont
			{White}}\ and\ \bibinfo {author} {\bibfnamefont {D.~J.}\ \bibnamefont
			{Scalapino}},\ }\bibfield  {title} {\bibinfo {title} {{Density Matrix
				Renormalization Group Study of the Striped Phase in the 2D $t-J$ Model}},\
	}\href {https://doi.org/10.1103/physrevlett.80.1272} {\bibfield  {journal}
		{\bibinfo  {journal} {Phys. Rev. Lett.}\ }\textbf {\bibinfo {volume} {80}},\
		\bibinfo {pages} {1272} (\bibinfo {year} {1998})}\BibitemShut {NoStop}%
	\bibitem [{\citenamefont {Yan}\ \emph {et~al.}(2011)\citenamefont {Yan},
		\citenamefont {Huse},\ and\ \citenamefont {White}}]{Yan2011}%
	\BibitemOpen
	\bibfield  {author} {\bibinfo {author} {\bibfnamefont {S.}~\bibnamefont
			{Yan}}, \bibinfo {author} {\bibfnamefont {D.~A.}\ \bibnamefont {Huse}},\ and\
		\bibinfo {author} {\bibfnamefont {S.~R.}\ \bibnamefont {White}},\ }\bibfield
	{title} {\bibinfo {title} {{Spin-Liquid Ground State of the $S = 1/2$ Kagome
				Heisenberg Antiferromagnet}},\ }\href
	{https://doi.org/10.1126/science.1201080} {\bibfield  {journal} {\bibinfo
			{journal} {Science}\ }\textbf {\bibinfo {volume} {332}},\ \bibinfo {pages}
		{1201080} (\bibinfo {year} {2011})}\BibitemShut {NoStop}%
	\bibitem [{\citenamefont {Corboz}\ and\ \citenamefont
		{Mila}(2013)}]{Corboz2013}%
	\BibitemOpen
	\bibfield  {author} {\bibinfo {author} {\bibfnamefont {P.}~\bibnamefont
			{Corboz}}\ and\ \bibinfo {author} {\bibfnamefont {F.}~\bibnamefont {Mila}},\
	}\bibfield  {title} {\bibinfo {title} {Tensor network study of the
			{S}hastry-{S}utherland model in zero magnetic field},\ }\href
	{https://doi.org/10.1103/physrevb.87.115144} {\bibfield  {journal} {\bibinfo
			{journal} {Phys. Rev. B}\ }\textbf {\bibinfo {volume} {87}},\ \bibinfo
		{pages} {115144} (\bibinfo {year} {2013})}\BibitemShut {NoStop}%
	\bibitem [{\citenamefont {Corboz}\ \emph {et~al.}(2014)\citenamefont {Corboz},
		\citenamefont {Rice},\ and\ \citenamefont {Troyer}}]{Corboz2014}%
	\BibitemOpen
	\bibfield  {author} {\bibinfo {author} {\bibfnamefont {P.}~\bibnamefont
			{Corboz}}, \bibinfo {author} {\bibfnamefont {T.~M.}\ \bibnamefont {Rice}},\
		and\ \bibinfo {author} {\bibfnamefont {M.}~\bibnamefont {Troyer}},\
	}\bibfield  {title} {\bibinfo {title} {{Competing States in the $t-J$-Model:
				Uniform $d$-Wave State versus Stripe State}},\ }\href
	{https://doi.org/10.1103/physrevlett.113.046402} {\bibfield  {journal}
		{\bibinfo  {journal} {Phys. Rev. Lett.}\ }\textbf {\bibinfo {volume} {113}},\
		\bibinfo {pages} {046402} (\bibinfo {year} {2014})}\BibitemShut {NoStop}%
	\bibitem [{\citenamefont {He}\ \emph {et~al.}(2017)\citenamefont {He},
		\citenamefont {Zaletel}, \citenamefont {Oshikawa},\ and\ \citenamefont
		{Pollmann}}]{He2017}%
	\BibitemOpen
	\bibfield  {author} {\bibinfo {author} {\bibfnamefont {Y.-C.}\ \bibnamefont
			{He}}, \bibinfo {author} {\bibfnamefont {M.~P.}\ \bibnamefont {Zaletel}},
		\bibinfo {author} {\bibfnamefont {M.}~\bibnamefont {Oshikawa}},\ and\
		\bibinfo {author} {\bibfnamefont {F.}~\bibnamefont {Pollmann}},\ }\bibfield
	{title} {\bibinfo {title} {Signatures of {D}irac cones in a {DMRG} study of
			the kagome {H}eisenberg model},\ }\href
	{https://doi.org/10.1103/physrevx.7.031020} {\bibfield  {journal} {\bibinfo
			{journal} {Phys. Rev. X}\ }\textbf {\bibinfo {volume} {7}},\ \bibinfo {pages}
		{031020} (\bibinfo {year} {2017})}\BibitemShut {NoStop}%
	\bibitem [{\citenamefont {Gohlke}\ \emph {et~al.}(2018)\citenamefont {Gohlke},
		\citenamefont {Wachtel}, \citenamefont {Yamaji}, \citenamefont {Pollmann},\
		and\ \citenamefont {Kim}}]{Gohlke2018}%
	\BibitemOpen
	\bibfield  {author} {\bibinfo {author} {\bibfnamefont {M.}~\bibnamefont
			{Gohlke}}, \bibinfo {author} {\bibfnamefont {G.}~\bibnamefont {Wachtel}},
		\bibinfo {author} {\bibfnamefont {Y.}~\bibnamefont {Yamaji}}, \bibinfo
		{author} {\bibfnamefont {F.}~\bibnamefont {Pollmann}},\ and\ \bibinfo
		{author} {\bibfnamefont {Y.~B.}\ \bibnamefont {Kim}},\ }\bibfield  {title}
	{\bibinfo {title} {Quantum spin liquid signatures in {K}itaev-like frustrated
			magnets},\ }\href {https://doi.org/10.1103/physrevb.97.075126} {\bibfield
		{journal} {\bibinfo  {journal} {Phys. Rev. B}\ }\textbf {\bibinfo {volume}
			{97}},\ \bibinfo {pages} {075126} (\bibinfo {year} {2018})}\BibitemShut
	{NoStop}%
	\bibitem [{\citenamefont {Pollmann}\ \emph {et~al.}(2010)\citenamefont
		{Pollmann}, \citenamefont {Turner}, \citenamefont {Berg},\ and\ \citenamefont
		{Oshikawa}}]{Pollmann2010}%
	\BibitemOpen
	\bibfield  {author} {\bibinfo {author} {\bibfnamefont {F.}~\bibnamefont
			{Pollmann}}, \bibinfo {author} {\bibfnamefont {A.~M.}\ \bibnamefont
			{Turner}}, \bibinfo {author} {\bibfnamefont {E.}~\bibnamefont {Berg}},\ and\
		\bibinfo {author} {\bibfnamefont {M.}~\bibnamefont {Oshikawa}},\ }\bibfield
	{title} {\bibinfo {title} {Entanglement spectrum of a topological phase in
			one dimension},\ }\href {https://doi.org/10.1103/physrevb.81.064439}
	{\bibfield  {journal} {\bibinfo  {journal} {Phys. Rev. B}\ }\textbf {\bibinfo
			{volume} {81}},\ \bibinfo {pages} {064439} (\bibinfo {year}
		{2010})}\BibitemShut {NoStop}%
	\bibitem [{\citenamefont {Schuch}\ \emph {et~al.}(2011)\citenamefont {Schuch},
		\citenamefont {P\'{e}rez-Garc\'{i}a},\ and\ \citenamefont
		{Cirac}}]{Schuch2011}%
	\BibitemOpen
	\bibfield  {author} {\bibinfo {author} {\bibfnamefont {N.}~\bibnamefont
			{Schuch}}, \bibinfo {author} {\bibfnamefont {D.}~\bibnamefont
			{P\'{e}rez-Garc\'{i}a}},\ and\ \bibinfo {author} {\bibfnamefont
			{I.}~\bibnamefont {Cirac}},\ }\bibfield  {title} {\bibinfo {title}
		{Classifying quantum phases using matrix product states and projected
			entangled pair states},\ }\href {https://doi.org/10.1103/physrevb.84.165139}
	{\bibfield  {journal} {\bibinfo  {journal} {Phys. Rev. B}\ }\textbf {\bibinfo
			{volume} {84}},\ \bibinfo {pages} {165139} (\bibinfo {year}
		{2011})}\BibitemShut {NoStop}%
	\bibitem [{\citenamefont {Williamson}\ \emph {et~al.}(2016)\citenamefont
		{Williamson}, \citenamefont {Bultinck}, \citenamefont {Mariën},
		\citenamefont {Şahinoğlu}, \citenamefont {Haegeman},\ and\ \citenamefont
		{Verstraete}}]{Williamson2016}%
	\BibitemOpen
	\bibfield  {author} {\bibinfo {author} {\bibfnamefont {D.~J.}\ \bibnamefont
			{Williamson}}, \bibinfo {author} {\bibfnamefont {N.}~\bibnamefont
			{Bultinck}}, \bibinfo {author} {\bibfnamefont {M.}~\bibnamefont {Mariën}},
		\bibinfo {author} {\bibfnamefont {M.~B.}\ \bibnamefont {Şahinoğlu}},
		\bibinfo {author} {\bibfnamefont {J.}~\bibnamefont {Haegeman}},\ and\
		\bibinfo {author} {\bibfnamefont {F.}~\bibnamefont {Verstraete}},\ }\bibfield
	{title} {\bibinfo {title} {Matrix product operators for symmetry-protected
			topological phases: Gauging and edge theories},\ }\href
	{https://doi.org/10.1103/physrevb.94.205150} {\bibfield  {journal} {\bibinfo
			{journal} {Phys. Rev. B}\ }\textbf {\bibinfo {volume} {94}},\ \bibinfo
		{pages} {205150} (\bibinfo {year} {2016})}\BibitemShut {NoStop}%
	\bibitem [{\citenamefont {Wahl}(2018)}]{Wahl2018}%
	\BibitemOpen
	\bibfield  {author} {\bibinfo {author} {\bibfnamefont {T.~B.}\ \bibnamefont
			{Wahl}},\ }\bibfield  {title} {\bibinfo {title} {Tensor networks demonstrate
			the robustness of localization and symmetry-protected topological phases},\
	}\href {https://doi.org/10.1103/physrevb.98.054204} {\bibfield  {journal}
		{\bibinfo  {journal} {Phys. Rev. B}\ }\textbf {\bibinfo {volume} {98}},\
		\bibinfo {pages} {054204} (\bibinfo {year} {2018})}\BibitemShut {NoStop}%
	\bibitem [{\citenamefont {Chan}\ and\ \citenamefont {Wahl}(2020)}]{Chan2020}%
	\BibitemOpen
	\bibfield  {author} {\bibinfo {author} {\bibfnamefont {A.}~\bibnamefont
			{Chan}}\ and\ \bibinfo {author} {\bibfnamefont {T.~B.}\ \bibnamefont
			{Wahl}},\ }\bibfield  {title} {\bibinfo {title} {Classification of
			symmetry-protected topological many-body localized phases in one dimension},\
	}\href {https://doi.org/10.1088/1361-648x/ab7f01} {\bibfield  {journal}
		{\bibinfo  {journal} {J. Phys. Cond. Mat.}\ }\textbf {\bibinfo {volume}
			{32}},\ \bibinfo {pages} {305601} (\bibinfo {year} {2020})}\BibitemShut
	{NoStop}%
	\bibitem [{\citenamefont {Li}\ \emph {et~al.}(2020)\citenamefont {Li},
		\citenamefont {Chan},\ and\ \citenamefont {Wahl}}]{Li2020}%
	\BibitemOpen
	\bibfield  {author} {\bibinfo {author} {\bibfnamefont {J.}~\bibnamefont
			{Li}}, \bibinfo {author} {\bibfnamefont {A.}~\bibnamefont {Chan}},\ and\
		\bibinfo {author} {\bibfnamefont {T.~B.}\ \bibnamefont {Wahl}},\ }\bibfield
	{title} {\bibinfo {title} {Classification of symmetry-protected topological
			phases in two-dimensional many-body localized systems},\ }\href
	{https://doi.org/10.1103/physrevb.102.014205} {\bibfield  {journal} {\bibinfo
			{journal} {Phys. Rev. B}\ }\textbf {\bibinfo {volume} {102}},\ \bibinfo
		{pages} {014205} (\bibinfo {year} {2020})}\BibitemShut {NoStop}%
	\bibitem [{\citenamefont {Qi}\ and\ \citenamefont {Zhang}(2011)}]{Rmp1}%
	\BibitemOpen
	\bibfield  {author} {\bibinfo {author} {\bibfnamefont {X.-L.}\ \bibnamefont
			{Qi}}\ and\ \bibinfo {author} {\bibfnamefont {S.-C.}\ \bibnamefont {Zhang}},\
	}\bibfield  {title} {\bibinfo {title} {Topological insulators and
			superconductors},\ }\href {https://doi.org/10.1103/RevModPhys.83.1057}
	{\bibfield  {journal} {\bibinfo  {journal} {Rev. Mod. Phys.}\ }\textbf
		{\bibinfo {volume} {83}},\ \bibinfo {pages} {1057} (\bibinfo {year}
		{2011})}\BibitemShut {NoStop}%
	\bibitem [{\citenamefont {Hasan}\ and\ \citenamefont {Kane}(2010)}]{Rmp2}%
	\BibitemOpen
	\bibfield  {author} {\bibinfo {author} {\bibfnamefont {M.~Z.}\ \bibnamefont
			{Hasan}}\ and\ \bibinfo {author} {\bibfnamefont {C.~L.}\ \bibnamefont
			{Kane}},\ }\bibfield  {title} {\bibinfo {title} {Colloquium},\ }\href
	{https://doi.org/10.1103/RevModPhys.82.3045} {\bibfield  {journal} {\bibinfo
			{journal} {Rev. Mod. Phys.}\ }\textbf {\bibinfo {volume} {82}},\ \bibinfo
		{pages} {3045} (\bibinfo {year} {2010})}\BibitemShut {NoStop}%
	\bibitem [{\citenamefont {Fu}(2011)}]{Clas1}%
	\BibitemOpen
	\bibfield  {author} {\bibinfo {author} {\bibfnamefont {L.}~\bibnamefont
			{Fu}},\ }\bibfield  {title} {\bibinfo {title} {{Topological Crystalline
				Insulators}},\ }\href {https://doi.org/10.1103/PhysRevLett.106.106802}
	{\bibfield  {journal} {\bibinfo  {journal} {Phys. Rev. Lett.}\ }\textbf
		{\bibinfo {volume} {106}},\ \bibinfo {pages} {106802} (\bibinfo {year}
		{2011})}\BibitemShut {NoStop}%
	\bibitem [{\citenamefont {Slager}\ \emph {et~al.}(2013)\citenamefont {Slager},
		\citenamefont {Mesaros}, \citenamefont {Juri{\v c}i{\'c}},\ and\
		\citenamefont {Zaanen}}]{Clas2}%
	\BibitemOpen
	\bibfield  {author} {\bibinfo {author} {\bibfnamefont {R.-J.}\ \bibnamefont
			{Slager}}, \bibinfo {author} {\bibfnamefont {A.}~\bibnamefont {Mesaros}},
		\bibinfo {author} {\bibfnamefont {V.}~\bibnamefont {Juri{\v c}i{\'c}}},\ and\
		\bibinfo {author} {\bibfnamefont {J.}~\bibnamefont {Zaanen}},\ }\bibfield
	{title} {\bibinfo {title} {The space group classification of topological
			band-insulators},\ }\href {http://dx.doi.org/10.1038/nphys2513} {\bibfield
		{journal} {\bibinfo  {journal} {Nat. Phys.}\ }\textbf {\bibinfo {volume}
			{9}},\ \bibinfo {pages} {98} (\bibinfo {year} {2013})}\BibitemShut {NoStop}%
	\bibitem [{\citenamefont {Po}\ \emph {et~al.}(2017)\citenamefont {Po},
		\citenamefont {Vishwanath},\ and\ \citenamefont {Watanabe}}]{clas4}%
	\BibitemOpen
	\bibfield  {author} {\bibinfo {author} {\bibfnamefont {H.~C.}\ \bibnamefont
			{Po}}, \bibinfo {author} {\bibfnamefont {A.}~\bibnamefont {Vishwanath}},\
		and\ \bibinfo {author} {\bibfnamefont {H.}~\bibnamefont {Watanabe}},\
	}\bibfield  {title} {\bibinfo {title} {Symmetry-based indicators of band
			topology in the 230 space groups},\ }\href
	{https://doi.org/10.1038/s41467-017-00133-2} {\bibfield  {journal} {\bibinfo
			{journal} {Nat. Commun.}\ }\textbf {\bibinfo {volume} {8}},\ \bibinfo {pages}
		{50} (\bibinfo {year} {2017})}\BibitemShut {NoStop}%
	\bibitem [{\citenamefont {Bradlyn}\ \emph {et~al.}(2017)\citenamefont
		{Bradlyn}, \citenamefont {Elcoro}, \citenamefont {Cano}, \citenamefont
		{Vergniory}, \citenamefont {Wang}, \citenamefont {Felser}, \citenamefont
		{Aroyo},\ and\ \citenamefont {Bernevig}}]{Clas5}%
	\BibitemOpen
	\bibfield  {author} {\bibinfo {author} {\bibfnamefont {B.}~\bibnamefont
			{Bradlyn}}, \bibinfo {author} {\bibfnamefont {L.}~\bibnamefont {Elcoro}},
		\bibinfo {author} {\bibfnamefont {J.}~\bibnamefont {Cano}}, \bibinfo {author}
		{\bibfnamefont {M.~G.}\ \bibnamefont {Vergniory}}, \bibinfo {author}
		{\bibfnamefont {Z.}~\bibnamefont {Wang}}, \bibinfo {author} {\bibfnamefont
			{C.}~\bibnamefont {Felser}}, \bibinfo {author} {\bibfnamefont {M.~I.}\
			\bibnamefont {Aroyo}},\ and\ \bibinfo {author} {\bibfnamefont {B.~A.}\
			\bibnamefont {Bernevig}},\ }\bibfield  {title} {\bibinfo {title} {Topological
			quantum chemistry},\ }\href {http://dx.doi.org/10.1038/nature23268}
	{\bibfield  {journal} {\bibinfo  {journal} {Nature}\ }\textbf {\bibinfo
			{volume} {547}},\ \bibinfo {pages} {298} (\bibinfo {year}
		{2017})}\BibitemShut {NoStop}%
	\bibitem [{\citenamefont {Slager}(2019)}]{Clas6}%
	\BibitemOpen
	\bibfield  {author} {\bibinfo {author} {\bibfnamefont {R.-J.}\ \bibnamefont
			{Slager}},\ }\bibfield  {title} {\bibinfo {title} {The translational side of
			topological band insulators},\ }\href
	{https://doi.org/https://doi.org/10.1016/j.jpcs.2018.01.023} {\bibfield
		{journal} {\bibinfo  {journal} {J. Phys. Chem. Solids}\ }\textbf {\bibinfo
			{volume} {128}},\ \bibinfo {pages} {24} (\bibinfo {year} {2019})}\BibitemShut
	{NoStop}%
	\bibitem [{\citenamefont {Bouhon}\ \emph
		{et~al.}(2020{\natexlab{a}})\citenamefont {Bouhon}, \citenamefont
		{Bzdu\v{s}ek},\ and\ \citenamefont {Slager}}]{Clas7}%
	\BibitemOpen
	\bibfield  {author} {\bibinfo {author} {\bibfnamefont {A.}~\bibnamefont
			{Bouhon}}, \bibinfo {author} {\bibfnamefont {T.}~\bibnamefont
			{Bzdu\v{s}ek}},\ and\ \bibinfo {author} {\bibfnamefont {R.-J.}\ \bibnamefont
			{Slager}},\ }\bibfield  {title} {\bibinfo {title} {Geometric approach to
			fragile topology beyond symmetry indicators},\ }\href
	{https://doi.org/10.1103/PhysRevB.102.115135} {\bibfield  {journal} {\bibinfo
			{journal} {Phys. Rev. B}\ }\textbf {\bibinfo {volume} {102}},\ \bibinfo
		{pages} {115135} (\bibinfo {year} {2020}{\natexlab{a}})}\BibitemShut
	{NoStop}%
	\bibitem [{\citenamefont {Shiozaki}\ and\ \citenamefont
		{Sato}(2014)}]{Shiozaki14}%
	\BibitemOpen
	\bibfield  {author} {\bibinfo {author} {\bibfnamefont {K.}~\bibnamefont
			{Shiozaki}}\ and\ \bibinfo {author} {\bibfnamefont {M.}~\bibnamefont
			{Sato}},\ }\bibfield  {title} {\bibinfo {title} {Topology of crystalline
			insulators and superconductors},\ }\href
	{https://doi.org/10.1103/PhysRevB.90.165114} {\bibfield  {journal} {\bibinfo
			{journal} {Phys. Rev. B}\ }\textbf {\bibinfo {volume} {90}},\ \bibinfo
		{pages} {165114} (\bibinfo {year} {2014})}\BibitemShut {NoStop}%
	\bibitem [{\citenamefont {Dubail}\ and\ \citenamefont
		{Read}(2015)}]{Dubail2015}%
	\BibitemOpen
	\bibfield  {author} {\bibinfo {author} {\bibfnamefont {J.}~\bibnamefont
			{Dubail}}\ and\ \bibinfo {author} {\bibfnamefont {N.}~\bibnamefont {Read}},\
	}\bibfield  {title} {\bibinfo {title} {Tensor network trial states for chiral
			topological phases in two dimensions and a no-go theorem in any dimension},\
	}\href {http://dx.doi.org/10.1103/PhysRevB.92.205307} {\bibfield  {journal}
		{\bibinfo  {journal} {Phys. Rev. B}\ }\textbf {\bibinfo {volume} {92}},\
		\bibinfo {pages} {205307} (\bibinfo {year} {2015})}\BibitemShut {NoStop}%
	\bibitem [{\citenamefont {Wahl}\ \emph {et~al.}(2013)\citenamefont {Wahl},
		\citenamefont {Tu}, \citenamefont {Schuch},\ and\ \citenamefont
		{Cirac}}]{Wahl2013}%
	\BibitemOpen
	\bibfield  {author} {\bibinfo {author} {\bibfnamefont {T.~B.}\ \bibnamefont
			{Wahl}}, \bibinfo {author} {\bibfnamefont {H.-H.}\ \bibnamefont {Tu}},
		\bibinfo {author} {\bibfnamefont {N.}~\bibnamefont {Schuch}},\ and\ \bibinfo
		{author} {\bibfnamefont {J.~I.}\ \bibnamefont {Cirac}},\ }\bibfield  {title}
	{\bibinfo {title} {{Projected Entangled-Pair States Can Describe Chiral
				Topological States}},\ }\href
	{http://dx.doi.org/10.1103/PhysRevLett.111.236805} {\bibfield  {journal}
		{\bibinfo  {journal} {Phys. Rev. Lett.}\ }\textbf {\bibinfo {volume} {111}},\
		\bibinfo {pages} {236805} (\bibinfo {year} {2013})}\BibitemShut {NoStop}%
	\bibitem [{\citenamefont {Yang}\ \emph {et~al.}(2015)\citenamefont {Yang},
		\citenamefont {Wahl}, \citenamefont {Tu}, \citenamefont {Schuch},\ and\
		\citenamefont {Cirac}}]{Yang2015}%
	\BibitemOpen
	\bibfield  {author} {\bibinfo {author} {\bibfnamefont {S.}~\bibnamefont
			{Yang}}, \bibinfo {author} {\bibfnamefont {T.~B.}\ \bibnamefont {Wahl}},
		\bibinfo {author} {\bibfnamefont {H.-H.}\ \bibnamefont {Tu}}, \bibinfo
		{author} {\bibfnamefont {N.}~\bibnamefont {Schuch}},\ and\ \bibinfo {author}
		{\bibfnamefont {J.~I.}\ \bibnamefont {Cirac}},\ }\bibfield  {title} {\bibinfo
		{title} {{Chiral Projected Entangled-Pair State with Topological Order}},\
	}\href {http://dx.doi.org/10.1103/PhysRevLett.114.106803} {\bibfield
		{journal} {\bibinfo  {journal} {Phys. Rev. Lett.}\ }\textbf {\bibinfo
			{volume} {114}},\ \bibinfo {pages} {106803} (\bibinfo {year}
		{2015})}\BibitemShut {NoStop}%
	\bibitem [{\citenamefont {Kitaev}(2009)}]{Kitaev}%
	\BibitemOpen
	\bibfield  {author} {\bibinfo {author} {\bibfnamefont {A.}~\bibnamefont
			{Kitaev}},\ }\bibfield  {title} {\bibinfo {title} {Periodic table for
			topological insulators and superconductors},\ }\href
	{https://doi.org/10.1063/1.3149495} {\bibfield  {journal} {\bibinfo
			{journal} {AIP Conf. Proc.}\ }\textbf {\bibinfo {volume} {1134}},\ \bibinfo
		{pages} {22} (\bibinfo {year} {2009})}\BibitemShut {NoStop}%
	\bibitem [{\citenamefont {Read}(2017)}]{Read2017}%
	\BibitemOpen
	\bibfield  {author} {\bibinfo {author} {\bibfnamefont {N.}~\bibnamefont
			{Read}},\ }\bibfield  {title} {\bibinfo {title} {Compactly supported
			{W}annier functions and algebraic ${K}$-theory},\ }\href
	{http://dx.doi.org/10.1103/PhysRevB.95.115309} {\bibfield  {journal}
		{\bibinfo  {journal} {Phys. Rev. B}\ }\textbf {\bibinfo {volume} {95}},\
		\bibinfo {pages} {115309} (\bibinfo {year} {2017})}\BibitemShut {NoStop}%
	\bibitem [{\citenamefont {T\"orm\"a}(2023)}]{tormaessay}%
	\BibitemOpen
	\bibfield  {author} {\bibinfo {author} {\bibfnamefont {P.}~\bibnamefont
			{T\"orm\"a}},\ }\bibfield  {title} {\bibinfo {title} {{Essay: Where Can
				Quantum Geometry Lead Us?}},\ }\href
	{https://doi.org/10.1103/PhysRevLett.131.240001} {\bibfield  {journal}
		{\bibinfo  {journal} {Phys. Rev. Lett.}\ }\textbf {\bibinfo {volume} {131}},\
		\bibinfo {pages} {240001} (\bibinfo {year} {2023})}\BibitemShut {NoStop}%
	\bibitem [{\citenamefont {Bouhon}\ \emph {et~al.}(2023)\citenamefont {Bouhon},
		\citenamefont {Timmel},\ and\ \citenamefont {Slager}}]{bouhon2023quantum}%
	\BibitemOpen
	\bibfield  {author} {\bibinfo {author} {\bibfnamefont {A.}~\bibnamefont
			{Bouhon}}, \bibinfo {author} {\bibfnamefont {A.}~\bibnamefont {Timmel}},\
		and\ \bibinfo {author} {\bibfnamefont {R.-J.}\ \bibnamefont {Slager}},\
	}\href {https://arxiv.org/abs/2303.02180} {\bibinfo {title} {Quantum geometry
			beyond projective single bands}} (\bibinfo {year} {2023})\BibitemShut
	{NoStop}%
	\bibitem [{\citenamefont {Provost}\ and\ \citenamefont
		{Vallee}(1980)}]{provost1980riemannian}%
	\BibitemOpen
	\bibfield  {author} {\bibinfo {author} {\bibfnamefont {J.}~\bibnamefont
			{Provost}}\ and\ \bibinfo {author} {\bibfnamefont {G.}~\bibnamefont
			{Vallee}},\ }\bibfield  {title} {\bibinfo {title} {Riemannian structure on
			manifolds of quantum states},\ }\href@noop {} {\bibfield  {journal} {\bibinfo
			{journal} {Commun. Math. Phys.}\ }\textbf {\bibinfo {volume} {76}},\
		\bibinfo {pages} {289} (\bibinfo {year} {1980})}\BibitemShut {NoStop}%
	\bibitem [{\citenamefont {Resta}(2011)}]{resta_2011_metric}%
	\BibitemOpen
	\bibfield  {author} {\bibinfo {author} {\bibfnamefont {R.}~\bibnamefont
			{Resta}},\ }\bibfield  {title} {\bibinfo {title} {The insulating state of
			matter: a geometrical theory},\ }\href
	{https://doi.org/10.1140/epjb/e2010-10874-4} {\bibfield  {journal} {\bibinfo
			{journal} {Euro. Phys. Jour. B}\ }\textbf {\bibinfo {volume} {79}},\ \bibinfo
		{pages} {121} (\bibinfo {year} {2011})}\BibitemShut {NoStop}%
	\bibitem [{\citenamefont {Verstraete}\ and\ \citenamefont
		{Cirac}(2004)}]{PEPS}%
	\BibitemOpen
	\bibfield  {author} {\bibinfo {author} {\bibfnamefont {F.}~\bibnamefont
			{Verstraete}}\ and\ \bibinfo {author} {\bibfnamefont {J.~I.}\ \bibnamefont
			{Cirac}},\ }\bibfield  {title} {\bibinfo {title} {Renormalization algorithms
			for quantum-many body systems in two and higher dimensions},\ }\href
	{https://arxiv.org/abs/cond-mat/0407066} {\bibfield  {journal} {\bibinfo
			{journal} {Preprint at https://arxiv.org/abs/cond-mat/0407066}\ } (\bibinfo
		{year} {2004})}\BibitemShut {NoStop}%
	\bibitem [{\citenamefont {Bouhon}\ \emph
		{et~al.}(2020{\natexlab{b}})\citenamefont {Bouhon}, \citenamefont {Wu},
		\citenamefont {Slager}, \citenamefont {Weng}, \citenamefont {Yazyev},\ and\
		\citenamefont {Bzdu{\v s}ek}}]{bouhon2019nonabelian}%
	\BibitemOpen
	\bibfield  {author} {\bibinfo {author} {\bibfnamefont {A.}~\bibnamefont
			{Bouhon}}, \bibinfo {author} {\bibfnamefont {Q.}~\bibnamefont {Wu}}, \bibinfo
		{author} {\bibfnamefont {R.-J.}\ \bibnamefont {Slager}}, \bibinfo {author}
		{\bibfnamefont {H.}~\bibnamefont {Weng}}, \bibinfo {author} {\bibfnamefont
			{O.~V.}\ \bibnamefont {Yazyev}},\ and\ \bibinfo {author} {\bibfnamefont
			{T.}~\bibnamefont {Bzdu{\v s}ek}},\ }\bibfield  {title} {\bibinfo {title}
		{Non-{A}belian reciprocal braiding of {W}eyl points and its manifestation in
			{Z}r{T}e},\ }\href {https://doi.org/10.1038/s41567-020-0967-9} {\bibfield
		{journal} {\bibinfo  {journal} {Nat. Phys.}\ }\textbf {\bibinfo {volume}
			{16}},\ \bibinfo {pages} {1137} (\bibinfo {year}
		{2020}{\natexlab{b}})}\BibitemShut {NoStop}%
	\bibitem [{\citenamefont {Ahn}\ \emph {et~al.}(2019)\citenamefont {Ahn},
		\citenamefont {Park},\ and\ \citenamefont {Yang}}]{BJY_nielsen}%
	\BibitemOpen
	\bibfield  {author} {\bibinfo {author} {\bibfnamefont {J.}~\bibnamefont
			{Ahn}}, \bibinfo {author} {\bibfnamefont {S.}~\bibnamefont {Park}},\ and\
		\bibinfo {author} {\bibfnamefont {B.-J.}\ \bibnamefont {Yang}},\ }\bibfield
	{title} {\bibinfo {title} {{Failure of Nielsen-Ninomiya Theorem and Fragile
				Topology in Two-Dimensional Systems with Space-Time Inversion Symmetry:
				Application to Twisted Bilayer Graphene at Magic Angle}},\ }\href
	{https://doi.org/10.1103/PhysRevX.9.021013} {\bibfield  {journal} {\bibinfo
			{journal} {Phys. Rev. X}\ }\textbf {\bibinfo {volume} {9}},\ \bibinfo {pages}
		{021013} (\bibinfo {year} {2019})}\BibitemShut {NoStop}%
	\bibitem [{\citenamefont {Bouhon}\ \emph {et~al.}(2019)\citenamefont {Bouhon},
		\citenamefont {Black-Schaffer},\ and\ \citenamefont
		{Slager}}]{Bouhon2018Wilson}%
	\BibitemOpen
	\bibfield  {author} {\bibinfo {author} {\bibfnamefont {A.}~\bibnamefont
			{Bouhon}}, \bibinfo {author} {\bibfnamefont {A.~M.}\ \bibnamefont
			{Black-Schaffer}},\ and\ \bibinfo {author} {\bibfnamefont {R.-J.}\
			\bibnamefont {Slager}},\ }\bibfield  {title} {\bibinfo {title} {Wilson loop
			approach to fragile topology of split elementary band representations and
			topological crystalline insulators with time-reversal symmetry},\ }\href
	{https://doi.org/10.1103/PhysRevB.100.195135} {\bibfield  {journal} {\bibinfo
			{journal} {Phys. Rev. B}\ }\textbf {\bibinfo {volume} {100}},\ \bibinfo
		{pages} {195135} (\bibinfo {year} {2019})}\BibitemShut {NoStop}%
	\bibitem [{\citenamefont {Wu}\ \emph {et~al.}(2019)\citenamefont {Wu},
		\citenamefont {Soluyanov},\ and\ \citenamefont
		{Bzdušek}}]{doi:10.1126/science.aau8740}%
	\BibitemOpen
	\bibfield  {author} {\bibinfo {author} {\bibfnamefont {Q.}~\bibnamefont
			{Wu}}, \bibinfo {author} {\bibfnamefont {A.~A.}\ \bibnamefont {Soluyanov}},\
		and\ \bibinfo {author} {\bibfnamefont {T.}~\bibnamefont {Bzdušek}},\
	}\bibfield  {title} {\bibinfo {title} {Non-{A}belian band topology in
			noninteracting metals},\ }\href {https://doi.org/10.1126/science.aau8740}
	{\bibfield  {journal} {\bibinfo  {journal} {Science}\ }\textbf {\bibinfo
			{volume} {365}},\ \bibinfo {pages} {1273} (\bibinfo {year}
		{2019})}\BibitemShut {NoStop}%
	\bibitem [{\citenamefont {Slager}\ \emph {et~al.}(2024)\citenamefont {Slager},
		\citenamefont {Bouhon},\ and\ \citenamefont {{\"U}nal}}]{slager2024floquet}%
	\BibitemOpen
	\bibfield  {author} {\bibinfo {author} {\bibfnamefont {R.-J.}\ \bibnamefont
			{Slager}}, \bibinfo {author} {\bibfnamefont {A.}~\bibnamefont {Bouhon}},\
		and\ \bibinfo {author} {\bibfnamefont {F.~N.}\ \bibnamefont {{\"U}nal}},\
	}\bibfield  {title} {\bibinfo {title} {Non-{A}belian {F}loquet braiding and
			anomalous {D}irac string phase in periodically driven systems},\ }\href
	{https://doi.org/10.1038/s41467-024-45302-2} {\bibfield  {journal} {\bibinfo
			{journal} {Nat Commun}\ }\textbf {\bibinfo {volume} {15}},\ \bibinfo {pages}
		{1144} (\bibinfo {year} {2024})}\BibitemShut {NoStop}%
	\bibitem [{\citenamefont {\"Unal}\ \emph {et~al.}(2020)\citenamefont {\"Unal},
		\citenamefont {Bouhon},\ and\ \citenamefont {Slager}}]{Unal_2020}%
	\BibitemOpen
	\bibfield  {author} {\bibinfo {author} {\bibfnamefont {F.~N.}\ \bibnamefont
			{\"Unal}}, \bibinfo {author} {\bibfnamefont {A.}~\bibnamefont {Bouhon}},\
		and\ \bibinfo {author} {\bibfnamefont {R.-J.}\ \bibnamefont {Slager}},\
	}\bibfield  {title} {\bibinfo {title} {{Topological Euler Class as a
				Dynamical Observable in Optical Lattices}},\ }\href
	{https://doi.org/10.1103/PhysRevLett.125.053601} {\bibfield  {journal}
		{\bibinfo  {journal} {Phys. Rev. Lett.}\ }\textbf {\bibinfo {volume} {125}},\
		\bibinfo {pages} {053601} (\bibinfo {year} {2020})}\BibitemShut {NoStop}%
	\bibitem [{\citenamefont {Zhao}\ \emph {et~al.}(2022)\citenamefont {Zhao},
		\citenamefont {Yang}, \citenamefont {Jiang}, \citenamefont {Mao},
		\citenamefont {Guo}, \citenamefont {Qiu}, \citenamefont {Wang}, \citenamefont
		{Yao}, \citenamefont {He}, \citenamefont {Zhou}, \citenamefont {Xu},\ and\
		\citenamefont {Duan}}]{Zhao_2022}%
	\BibitemOpen
	\bibfield  {author} {\bibinfo {author} {\bibfnamefont {W.}~\bibnamefont
			{Zhao}}, \bibinfo {author} {\bibfnamefont {Y.-B.}\ \bibnamefont {Yang}},
		\bibinfo {author} {\bibfnamefont {Y.}~\bibnamefont {Jiang}}, \bibinfo
		{author} {\bibfnamefont {Z.}~\bibnamefont {Mao}}, \bibinfo {author}
		{\bibfnamefont {W.}~\bibnamefont {Guo}}, \bibinfo {author} {\bibfnamefont
			{L.}~\bibnamefont {Qiu}}, \bibinfo {author} {\bibfnamefont {G.}~\bibnamefont
			{Wang}}, \bibinfo {author} {\bibfnamefont {L.}~\bibnamefont {Yao}}, \bibinfo
		{author} {\bibfnamefont {L.}~\bibnamefont {He}}, \bibinfo {author}
		{\bibfnamefont {Z.}~\bibnamefont {Zhou}}, \bibinfo {author} {\bibfnamefont
			{Y.}~\bibnamefont {Xu}},\ and\ \bibinfo {author} {\bibfnamefont
			{L.}~\bibnamefont {Duan}},\ }\bibfield  {title} {\bibinfo {title} {Quantum
			simulation for topological {E}uler insulators},\ }\href
	{https://doi.org/10.1038/s42005-022-01001-2} {\bibfield  {journal} {\bibinfo
			{journal} {Commun. Phys.}\ }\textbf {\bibinfo {volume} {5}},\ \bibinfo
		{pages} {223} (\bibinfo {year} {2022})}\BibitemShut {NoStop}%
	\bibitem [{\citenamefont {Peng}\ \emph
		{et~al.}(2022{\natexlab{a}})\citenamefont {Peng}, \citenamefont {Bouhon},
		\citenamefont {Monserrat},\ and\ \citenamefont {Slager}}]{Peng2021}%
	\BibitemOpen
	\bibfield  {author} {\bibinfo {author} {\bibfnamefont {B.}~\bibnamefont
			{Peng}}, \bibinfo {author} {\bibfnamefont {A.}~\bibnamefont {Bouhon}},
		\bibinfo {author} {\bibfnamefont {B.}~\bibnamefont {Monserrat}},\ and\
		\bibinfo {author} {\bibfnamefont {R.-J.}\ \bibnamefont {Slager}},\ }\bibfield
	{title} {\bibinfo {title} {Phonons as a platform for non-{A}belian braiding
			and its manifestation in layered silicates},\ }\href
	{https://doi.org/10.1038/s41467-022-28046-9} {\bibfield  {journal} {\bibinfo
			{journal} {Nat. Commun.}\ }\textbf {\bibinfo {volume} {13}},\ \bibinfo
		{pages} {423} (\bibinfo {year} {2022}{\natexlab{a}})}\BibitemShut {NoStop}%
	\bibitem [{\citenamefont {Peng}\ \emph
		{et~al.}(2022{\natexlab{b}})\citenamefont {Peng}, \citenamefont {Bouhon},
		\citenamefont {Slager},\ and\ \citenamefont {Monserrat}}]{Peng2022Multi}%
	\BibitemOpen
	\bibfield  {author} {\bibinfo {author} {\bibfnamefont {B.}~\bibnamefont
			{Peng}}, \bibinfo {author} {\bibfnamefont {A.}~\bibnamefont {Bouhon}},
		\bibinfo {author} {\bibfnamefont {R.-J.}\ \bibnamefont {Slager}},\ and\
		\bibinfo {author} {\bibfnamefont {B.}~\bibnamefont {Monserrat}},\ }\bibfield
	{title} {\bibinfo {title} {Multigap topology and non-{A}belian braiding of
			phonons from first principles},\ }\href
	{https://doi.org/10.1103/PhysRevB.105.085115} {\bibfield  {journal} {\bibinfo
			{journal} {Phys. Rev. B}\ }\textbf {\bibinfo {volume} {105}},\ \bibinfo
		{pages} {085115} (\bibinfo {year} {2022}{\natexlab{b}})}\BibitemShut
	{NoStop}%
	\bibitem [{\citenamefont {Bouhon}\ \emph {et~al.}(2021)\citenamefont {Bouhon},
		\citenamefont {Lange},\ and\ \citenamefont {Slager}}]{magnetic}%
	\BibitemOpen
	\bibfield  {author} {\bibinfo {author} {\bibfnamefont {A.}~\bibnamefont
			{Bouhon}}, \bibinfo {author} {\bibfnamefont {G.~F.}\ \bibnamefont {Lange}},\
		and\ \bibinfo {author} {\bibfnamefont {R.-J.}\ \bibnamefont {Slager}},\
	}\bibfield  {title} {\bibinfo {title} {Topological correspondence between
			magnetic space group representations and subdimensions},\ }\href
	{https://doi.org/10.1103/PhysRevB.103.245127} {\bibfield  {journal} {\bibinfo
			{journal} {Phys. Rev. B}\ }\textbf {\bibinfo {volume} {103}},\ \bibinfo
		{pages} {245127} (\bibinfo {year} {2021})}\BibitemShut {NoStop}%
	\bibitem [{\citenamefont {Lee}\ \emph {et~al.}(2024)\citenamefont {Lee},
		\citenamefont {Qian},\ and\ \citenamefont
		{Yang}}]{lee2024eulerbandtopologyspinorbit}%
	\BibitemOpen
	\bibfield  {author} {\bibinfo {author} {\bibfnamefont {S.~H.}\ \bibnamefont
			{Lee}}, \bibinfo {author} {\bibfnamefont {Y.}~\bibnamefont {Qian}},\ and\
		\bibinfo {author} {\bibfnamefont {B.-J.}\ \bibnamefont {Yang}},\ }\bibfield
	{title} {\bibinfo {title} {{E}uler band topology in spin-orbit coupled
			magnetic systems},\ }\href {https://arxiv.org/abs/2404.16383} {\bibfield
		{journal} {\bibinfo  {journal} {Preprint at
				https://arxiv.org/abs/2404.16383}\ } (\bibinfo {year} {2024})}\BibitemShut
	{NoStop}%
	\bibitem [{\citenamefont {Jiang}\ \emph {et~al.}(2021)\citenamefont {Jiang},
		\citenamefont {Bouhon}, \citenamefont {Lin}, \citenamefont {Zhou},
		\citenamefont {Hou}, \citenamefont {Li}, \citenamefont {Slager},\ and\
		\citenamefont {Jiang}}]{Jiang_2021}%
	\BibitemOpen
	\bibfield  {author} {\bibinfo {author} {\bibfnamefont {B.}~\bibnamefont
			{Jiang}}, \bibinfo {author} {\bibfnamefont {A.}~\bibnamefont {Bouhon}},
		\bibinfo {author} {\bibfnamefont {Z.-K.}\ \bibnamefont {Lin}}, \bibinfo
		{author} {\bibfnamefont {X.}~\bibnamefont {Zhou}}, \bibinfo {author}
		{\bibfnamefont {B.}~\bibnamefont {Hou}}, \bibinfo {author} {\bibfnamefont
			{F.}~\bibnamefont {Li}}, \bibinfo {author} {\bibfnamefont {R.-J.}\
			\bibnamefont {Slager}},\ and\ \bibinfo {author} {\bibfnamefont {J.-H.}\
			\bibnamefont {Jiang}},\ }\bibfield  {title} {\bibinfo {title} {Experimental
			observation of non-{A}belian topological acoustic semimetals and their phase
			transitions},\ }\href {https://doi.org/10.1038/s41567-021-01340-x} {\bibfield
		{journal} {\bibinfo  {journal} {Nat. Phys.}\ }\textbf {\bibinfo {volume}
			{17}},\ \bibinfo {pages} {1239} (\bibinfo {year} {2021})}\BibitemShut
	{NoStop}%
	\bibitem [{\citenamefont {Jiang}\ \emph {et~al.}(2024)\citenamefont {Jiang},
		\citenamefont {Bouhon}, \citenamefont {Wu}, \citenamefont {Kong},
		\citenamefont {Lin}, \citenamefont {Slager},\ and\ \citenamefont
		{Jiang}}]{JIANG2024}%
	\BibitemOpen
	\bibfield  {author} {\bibinfo {author} {\bibfnamefont {B.}~\bibnamefont
			{Jiang}}, \bibinfo {author} {\bibfnamefont {A.}~\bibnamefont {Bouhon}},
		\bibinfo {author} {\bibfnamefont {S.-Q.}\ \bibnamefont {Wu}}, \bibinfo
		{author} {\bibfnamefont {Z.-L.}\ \bibnamefont {Kong}}, \bibinfo {author}
		{\bibfnamefont {Z.-K.}\ \bibnamefont {Lin}}, \bibinfo {author} {\bibfnamefont
			{R.-J.}\ \bibnamefont {Slager}},\ and\ \bibinfo {author} {\bibfnamefont
			{J.-H.}\ \bibnamefont {Jiang}},\ }\bibfield  {title} {\bibinfo {title}
		{Observation of an acoustic topological {E}uler insulator with meronic
			waves},\ }\href {https://doi.org/https://doi.org/10.1016/j.scib.2024.04.009}
	{\bibfield  {journal} {\bibinfo  {journal} {Science Bulletin}\ }\textbf
		{\bibinfo {volume} {69}},\ \bibinfo {pages} {1653} (\bibinfo {year}
		{2024})}\BibitemShut {NoStop}%
	\bibitem [{\citenamefont {Guo}\ \emph {et~al.}(2021)\citenamefont {Guo},
		\citenamefont {Jiang}, \citenamefont {Zhang}, \citenamefont {Zhang},
		\citenamefont {Zhang}, \citenamefont {Yang}, \citenamefont {Zhang},\ and\
		\citenamefont {Chan}}]{Guo1Dexp}%
	\BibitemOpen
	\bibfield  {author} {\bibinfo {author} {\bibfnamefont {Q.}~\bibnamefont
			{Guo}}, \bibinfo {author} {\bibfnamefont {T.}~\bibnamefont {Jiang}}, \bibinfo
		{author} {\bibfnamefont {R.-Y.}\ \bibnamefont {Zhang}}, \bibinfo {author}
		{\bibfnamefont {L.}~\bibnamefont {Zhang}}, \bibinfo {author} {\bibfnamefont
			{Z.-Q.}\ \bibnamefont {Zhang}}, \bibinfo {author} {\bibfnamefont
			{B.}~\bibnamefont {Yang}}, \bibinfo {author} {\bibfnamefont {S.}~\bibnamefont
			{Zhang}},\ and\ \bibinfo {author} {\bibfnamefont {C.~T.}\ \bibnamefont
			{Chan}},\ }\bibfield  {title} {\bibinfo {title} {Experimental observation of
			non-{A}belian topological charges and edge states},\ }\href
	{https://doi.org/10.1038/s41586-021-03521-3} {\bibfield  {journal} {\bibinfo
			{journal} {Nature}\ }\textbf {\bibinfo {volume} {594}},\ \bibinfo {pages}
		{195} (\bibinfo {year} {2021})}\BibitemShut {NoStop}%
	\bibitem [{\citenamefont {Kitaev}(2001)}]{Kitaev2001}%
	\BibitemOpen
	\bibfield  {author} {\bibinfo {author} {\bibfnamefont {A.~Y.}\ \bibnamefont
			{Kitaev}},\ }\bibfield  {title} {\bibinfo {title} {Unpaired {M}ajorana
			fermions in quantum wires},\ }\href
	{https://doi.org/10.1070/1063-7869/44/10s/s29} {\bibfield  {journal}
		{\bibinfo  {journal} {Physics-Uspekhi}\ }\textbf {\bibinfo {volume} {44}},\
		\bibinfo {pages} {131} (\bibinfo {year} {2001})}\BibitemShut {NoStop}%
	\bibitem [{\citenamefont {Hackenbroich}\ \emph {et~al.}(2020)\citenamefont
		{Hackenbroich}, \citenamefont {Bernevig}, \citenamefont {Schuch},\ and\
		\citenamefont {Regnault}}]{Hackenbroich2020}%
	\BibitemOpen
	\bibfield  {author} {\bibinfo {author} {\bibfnamefont {A.}~\bibnamefont
			{Hackenbroich}}, \bibinfo {author} {\bibfnamefont {B.~A.}\ \bibnamefont
			{Bernevig}}, \bibinfo {author} {\bibfnamefont {N.}~\bibnamefont {Schuch}},\
		and\ \bibinfo {author} {\bibfnamefont {N.}~\bibnamefont {Regnault}},\
	}\bibfield  {title} {\bibinfo {title} {Fermionic tensor networks for
			higher-order topological insulators from charge pumping},\ }\href
	{https://doi.org/10.1103/physrevb.101.115134} {\bibfield  {journal} {\bibinfo
			{journal} {Phys. Rev. B}\ }\textbf {\bibinfo {volume} {101}},\ \bibinfo
		{pages} {115134} (\bibinfo {year} {2020})}\BibitemShut {NoStop}%
	\bibitem [{\citenamefont {Xie}\ \emph {et~al.}(2014)\citenamefont {Xie},
		\citenamefont {Chen}, \citenamefont {Yu}, \citenamefont {Kong}, \citenamefont
		{Normand},\ and\ \citenamefont {Xiang}}]{PESS}%
	\BibitemOpen
	\bibfield  {author} {\bibinfo {author} {\bibfnamefont {Z.}~\bibnamefont
			{Xie}}, \bibinfo {author} {\bibfnamefont {J.}~\bibnamefont {Chen}}, \bibinfo
		{author} {\bibfnamefont {J.}~\bibnamefont {Yu}}, \bibinfo {author}
		{\bibfnamefont {X.}~\bibnamefont {Kong}}, \bibinfo {author} {\bibfnamefont
			{B.}~\bibnamefont {Normand}},\ and\ \bibinfo {author} {\bibfnamefont
			{T.}~\bibnamefont {Xiang}},\ }\bibfield  {title} {\bibinfo {title} {{Tensor
				Renormalization of Quantum Many-Body Systems Using Projected Entangled
				Simplex States}},\ }\href {https://doi.org/10.1103/physrevx.4.011025}
	{\bibfield  {journal} {\bibinfo  {journal} {Phys. Rev. X}\ }\textbf {\bibinfo
			{volume} {4}},\ \bibinfo {pages} {011025} (\bibinfo {year}
		{2014})}\BibitemShut {NoStop}%
	\bibitem [{\citenamefont {Kraus}\ \emph {et~al.}(2010)\citenamefont {Kraus},
		\citenamefont {Schuch}, \citenamefont {Verstraete},\ and\ \citenamefont
		{Cirac}}]{Kraus2010}%
	\BibitemOpen
	\bibfield  {author} {\bibinfo {author} {\bibfnamefont {C.~V.}\ \bibnamefont
			{Kraus}}, \bibinfo {author} {\bibfnamefont {N.}~\bibnamefont {Schuch}},
		\bibinfo {author} {\bibfnamefont {F.}~\bibnamefont {Verstraete}},\ and\
		\bibinfo {author} {\bibfnamefont {J.~I.}\ \bibnamefont {Cirac}},\ }\bibfield
	{title} {\bibinfo {title} {Fermionic projected entangled pair states},\
	}\href {https://doi.org/10.1103/physreva.81.052338} {\bibfield  {journal}
		{\bibinfo  {journal} {Phys. Rev. A}\ }\textbf {\bibinfo {volume} {81}},\
		\bibinfo {pages} {052338} (\bibinfo {year} {2010})}\BibitemShut {NoStop}%
	\bibitem [{\citenamefont {D\"{u}r}\ \emph {et~al.}(2000)\citenamefont
		{D\"{u}r}, \citenamefont {Vidal},\ and\ \citenamefont {Cirac}}]{Duer2000}%
	\BibitemOpen
	\bibfield  {author} {\bibinfo {author} {\bibfnamefont {W.}~\bibnamefont
			{D\"{u}r}}, \bibinfo {author} {\bibfnamefont {G.}~\bibnamefont {Vidal}},\
		and\ \bibinfo {author} {\bibfnamefont {J.~I.}\ \bibnamefont {Cirac}},\
	}\bibfield  {title} {\bibinfo {title} {Three qubits can be entangled in two
			inequivalent ways},\ }\href {https://doi.org/10.1103/physreva.62.062314}
	{\bibfield  {journal} {\bibinfo  {journal} {Phys. Rev. A}\ }\textbf {\bibinfo
			{volume} {62}},\ \bibinfo {pages} {062314} (\bibinfo {year}
		{2000})}\BibitemShut {NoStop}%
	\bibitem [{\citenamefont {Wahl}\ \emph {et~al.}(2014)\citenamefont {Wahl},
		\citenamefont {Haßler}, \citenamefont {Tu}, \citenamefont {Cirac},\ and\
		\citenamefont {Schuch}}]{Wahl2014}%
	\BibitemOpen
	\bibfield  {author} {\bibinfo {author} {\bibfnamefont {T.~B.}\ \bibnamefont
			{Wahl}}, \bibinfo {author} {\bibfnamefont {S.~T.}\ \bibnamefont {Haßler}},
		\bibinfo {author} {\bibfnamefont {H.-H.}\ \bibnamefont {Tu}}, \bibinfo
		{author} {\bibfnamefont {J.~I.}\ \bibnamefont {Cirac}},\ and\ \bibinfo
		{author} {\bibfnamefont {N.}~\bibnamefont {Schuch}},\ }\bibfield  {title}
	{\bibinfo {title} {Symmetries and boundary theories for chiral projected
			entangled pair states},\ }\href
	{http://dx.doi.org/10.1103/PhysRevB.90.115133} {\bibfield  {journal}
		{\bibinfo  {journal} {Phys. Rev. B}\ }\textbf {\bibinfo {volume} {90}},\
		\bibinfo {pages} {115133} (\bibinfo {year} {2014})}\BibitemShut {NoStop}%
	\bibitem [{\citenamefont {Chen}\ \emph {et~al.}(2013)\citenamefont {Chen},
		\citenamefont {Gu}, \citenamefont {Liu},\ and\ \citenamefont
		{Wen}}]{Chen2013}%
	\BibitemOpen
	\bibfield  {author} {\bibinfo {author} {\bibfnamefont {X.}~\bibnamefont
			{Chen}}, \bibinfo {author} {\bibfnamefont {Z.-C.}\ \bibnamefont {Gu}},
		\bibinfo {author} {\bibfnamefont {Z.-X.}\ \bibnamefont {Liu}},\ and\ \bibinfo
		{author} {\bibfnamefont {X.-G.}\ \bibnamefont {Wen}},\ }\bibfield  {title}
	{\bibinfo {title} {Symmetry protected topological orders and the group
			cohomology of their symmetry group},\ }\href
	{https://doi.org/10.1103/physrevb.87.155114} {\bibfield  {journal} {\bibinfo
			{journal} {Phys. Rev. B}\ }\textbf {\bibinfo {volume} {87}},\ \bibinfo
		{pages} {155114} (\bibinfo {year} {2013})}\BibitemShut {NoStop}%
	\bibitem [{\citenamefont {Chen}\ \emph {et~al.}(2010)\citenamefont {Chen},
		\citenamefont {Gu},\ and\ \citenamefont {Wen}}]{Chen2010}%
	\BibitemOpen
	\bibfield  {author} {\bibinfo {author} {\bibfnamefont {X.}~\bibnamefont
			{Chen}}, \bibinfo {author} {\bibfnamefont {Z.-C.}\ \bibnamefont {Gu}},\ and\
		\bibinfo {author} {\bibfnamefont {X.-G.}\ \bibnamefont {Wen}},\ }\bibfield
	{title} {\bibinfo {title} {Local unitary transformation, long-range quantum
			entanglement, wave function renormalization, and topological order},\ }\href
	{https://doi.org/10.1103/PhysRevB.82.155138} {\bibfield  {journal} {\bibinfo
			{journal} {Phys. Rev. B}\ }\textbf {\bibinfo {volume} {82}},\ \bibinfo
		{pages} {155138} (\bibinfo {year} {2010})}\BibitemShut {NoStop}%
	\bibitem [{\citenamefont {Tu}(2013)}]{Tu2013}%
	\BibitemOpen
	\bibfield  {author} {\bibinfo {author} {\bibfnamefont {H.-H.}\ \bibnamefont
			{Tu}},\ }\bibfield  {title} {\bibinfo {title} {{Projected BCS states and spin
				Hamiltonians for the $SO(n)_1$ Wess-Zumino-Witten model}},\ }\href
	{https://doi.org/10.1103/physrevb.87.041103} {\bibfield  {journal} {\bibinfo
			{journal} {Phys. Rev. B}\ }\textbf {\bibinfo {volume} {87}},\ \bibinfo
		{pages} {041103} (\bibinfo {year} {2013})}\BibitemShut {NoStop}%
	\bibitem [{\citenamefont {Hauschild}\ and\ \citenamefont
		{Pollmann}(2018)}]{tenpy}%
	\BibitemOpen
	\bibfield  {author} {\bibinfo {author} {\bibfnamefont {J.}~\bibnamefont
			{Hauschild}}\ and\ \bibinfo {author} {\bibfnamefont {F.}~\bibnamefont
			{Pollmann}},\ }\bibfield  {title} {\bibinfo {title} {{Efficient numerical
				simulations with Tensor Networks: Tensor Network Python (TeNPy)}},\ }\href
	{https://doi.org/10.21468/SciPostPhysLectNotes.5} {\bibfield  {journal}
		{\bibinfo  {journal} {SciPost Phys. Lect. Notes}\ ,\ \bibinfo {pages} {5}}
		(\bibinfo {year} {2018})},\ \bibinfo {note} {code available from
		\url{https://github.com/tenpy/tenpy}}\BibitemShut {NoStop}%
	\bibitem [{\citenamefont {Bouhon}\ and\ \citenamefont
		{Slager}(2022)}]{bouhon2022multigap}%
	\BibitemOpen
	\bibfield  {author} {\bibinfo {author} {\bibfnamefont {A.}~\bibnamefont
			{Bouhon}}\ and\ \bibinfo {author} {\bibfnamefont {R.-J.}\ \bibnamefont
			{Slager}},\ }\bibfield  {title} {\bibinfo {title} {Multi-gap topological
			conversion of {E}uler class via band-node braiding: minimal models,
			${PT}$-linked nodal rings, and chiral heirs},\ }\href
	{https://arxiv.org/abs/2203.16741} {\bibfield  {journal} {\bibinfo  {journal}
			{Preprint at https://arxiv.org/abs/2203.16741}\ } (\bibinfo {year}
		{2022})}\BibitemShut {NoStop}%
	\bibitem [{\citenamefont {Peschel}(2003)}]{Peschel2003}%
	\BibitemOpen
	\bibfield  {author} {\bibinfo {author} {\bibfnamefont {I.}~\bibnamefont
			{Peschel}},\ }\bibfield  {title} {\bibinfo {title} {Calculation of reduced
			density matrices from correlation functions},\ }\href
	{https://doi.org/10.1088/0305-4470/36/14/101} {\bibfield  {journal} {\bibinfo
			{journal} {J. Phys. A Math. Gen.}\ }\textbf {\bibinfo {volume} {36}},\
		\bibinfo {pages} {L205} (\bibinfo {year} {2003})}\BibitemShut {NoStop}%
	\bibitem [{\citenamefont {Takahashi}\ and\ \citenamefont
		{Ozawa}(2023)}]{Takahashi_2023}%
	\BibitemOpen
	\bibfield  {author} {\bibinfo {author} {\bibfnamefont {R.}~\bibnamefont
			{Takahashi}}\ and\ \bibinfo {author} {\bibfnamefont {T.}~\bibnamefont
			{Ozawa}},\ }\bibfield  {title} {\bibinfo {title} {Bulk-edge correspondence of
			{S}tiefel-{W}hitney and {E}uler insulators through the entanglement spectrum
			and cutting procedure},\ }\href {https://doi.org/10.1103/physrevb.108.075129}
	{\bibfield  {journal} {\bibinfo  {journal} {Phys. Rev. B}\ }\textbf {\bibinfo
			{volume} {108}},\ \bibinfo {pages} {075129} (\bibinfo {year}
		{2023})}\BibitemShut {NoStop}%
\end{thebibliography}

\begin{thebibliography}{10}%
	\makeatletter
	\providecommand \@ifxundefined [1]{%
		\@ifx{#1\undefined}
	}%
	\providecommand \@ifnum [1]{%
		\ifnum #1\expandafter \@firstoftwo
		\else \expandafter \@secondoftwo
		\fi
	}%
	\providecommand \@ifx [1]{%
		\ifx #1\expandafter \@firstoftwo
		\else \expandafter \@secondoftwo
		\fi
	}%
	\providecommand \natexlab [1]{#1}%
	\providecommand \enquote  [1]{``#1''}%
	\providecommand \bibnamefont  [1]{#1}%
	\providecommand \bibfnamefont [1]{#1}%
	\providecommand \citenamefont [1]{#1}%
	\providecommand \href@noop [0]{\@secondoftwo}%
	\providecommand \href [0]{\begingroup \@sanitize@url \@href}%
	\providecommand \@href[1]{\@@startlink{#1}\@@href}%
	\providecommand \@@href[1]{\endgroup#1\@@endlink}%
	\providecommand \@sanitize@url [0]{\catcode `\\12\catcode `\$12\catcode
		`\&12\catcode `\#12\catcode `\^12\catcode `\_12\catcode `\%12\relax}%
	\providecommand \@@startlink[1]{}%
	\providecommand \@@endlink[0]{}%
	\providecommand \url  [0]{\begingroup\@sanitize@url \@url }%
	\providecommand \@url [1]{\endgroup\@href {#1}{\urlprefix }}%
	\providecommand \urlprefix  [0]{URL }%
	\providecommand \Eprint [0]{\href }%
	\providecommand \doibase [0]{https://doi.org/}%
	\providecommand \selectlanguage [0]{\@gobble}%
	\providecommand \bibinfo  [0]{\@secondoftwo}%
	\providecommand \bibfield  [0]{\@secondoftwo}%
	\providecommand \translation [1]{[#1]}%
	\providecommand \BibitemOpen [0]{}%
	\providecommand \bibitemStop [0]{}%
	\providecommand \bibitemNoStop [0]{.\EOS\space}%
	\providecommand \EOS [0]{\spacefactor3000\relax}%
	\providecommand \BibitemShut  [1]{\csname bibitem#1\endcsname}%
	\let\auto@bib@innerbib\@empty
	%</preamble>
	\bibitem [{\citenamefont {Provost}\ and\ \citenamefont
		{Vallee}(1980)}]{provost1980riemannian}%
	\BibitemOpen
	\bibfield  {author} {\bibinfo {author} {\bibfnamefont {J.}~\bibnamefont
			{Provost}}\ and\ \bibinfo {author} {\bibfnamefont {G.}~\bibnamefont
			{Vallee}},\ }\href@noop {} {\bibfield  {journal} {\bibinfo  {journal}
			{Commun. Math. Phys.}\ }\textbf {\bibinfo {volume} {76}},\ \bibinfo {pages}
		{289} (\bibinfo {year} {1980})}\BibitemShut {NoStop}%
	\bibitem [{\citenamefont {Bouhon}\ \emph {et~al.}(2023)\citenamefont {Bouhon},
		\citenamefont {Timmel},\ and\ \citenamefont {Slager}}]{bouhon2023quantum}%
	\BibitemOpen
	\bibfield  {author} {\bibinfo {author} {\bibfnamefont {A.}~\bibnamefont
			{Bouhon}}, \bibinfo {author} {\bibfnamefont {A.}~\bibnamefont {Timmel}},\
		and\ \bibinfo {author} {\bibfnamefont {R.-J.}\ \bibnamefont {Slager}},\
	}\href {https://arxiv.org/abs/2303.02180} {\bibfield  {journal} {\bibinfo
			{journal} {arxiv:2303.02180}\ } (\bibinfo {year} {2023})}\BibitemShut
	{NoStop}%
	\bibitem [{\citenamefont {T\"orm\"a}(2023)}]{tormaessay}%
	\BibitemOpen
	\bibfield  {author} {\bibinfo {author} {\bibfnamefont {P.}~\bibnamefont
			{T\"orm\"a}},\ }\href {https://doi.org/10.1103/PhysRevLett.131.240001}
	{\bibfield  {journal} {\bibinfo  {journal} {Phys. Rev. Lett.}\ }\textbf
		{\bibinfo {volume} {131}},\ \bibinfo {pages} {240001} (\bibinfo {year}
		{2023})}\BibitemShut {NoStop}%
	\bibitem [{\citenamefont {Kwon}\ and\ \citenamefont
		{Yang}(2024)}]{kwon2024quantum}%
	\BibitemOpen
	\bibfield  {author} {\bibinfo {author} {\bibfnamefont {S.}~\bibnamefont
			{Kwon}}\ and\ \bibinfo {author} {\bibfnamefont {B.-J.}\ \bibnamefont
			{Yang}},\ }\href {https://doi.org/10.1103/PhysRevB.109.L161111} {\bibfield
		{journal} {\bibinfo  {journal} {Phys. Rev. B}\ }\textbf {\bibinfo {volume}
			{109}},\ \bibinfo {pages} {L161111} (\bibinfo {year} {2024})}\BibitemShut
	{NoStop}%
	\bibitem [{\citenamefont {Braunstein}\ and\ \citenamefont
		{Caves}(1994)}]{PhysRevLett.72.3439}%
	\BibitemOpen
	\bibfield  {author} {\bibinfo {author} {\bibfnamefont {S.~L.}\ \bibnamefont
			{Braunstein}}\ and\ \bibinfo {author} {\bibfnamefont {C.~M.}\ \bibnamefont
			{Caves}},\ }\href {https://doi.org/10.1103/PhysRevLett.72.3439} {\bibfield
		{journal} {\bibinfo  {journal} {Phys. Rev. Lett.}\ }\textbf {\bibinfo
			{volume} {72}},\ \bibinfo {pages} {3439} (\bibinfo {year}
		{1994})}\BibitemShut {NoStop}%
	\bibitem [{\citenamefont {Liu}\ \emph {et~al.}(2019)\citenamefont {Liu},
		\citenamefont {Yuan}, \citenamefont {Lu},\ and\ \citenamefont
		{Wang}}]{Liu_2020}%
	\BibitemOpen
	\bibfield  {author} {\bibinfo {author} {\bibfnamefont {J.}~\bibnamefont
			{Liu}}, \bibinfo {author} {\bibfnamefont {H.}~\bibnamefont {Yuan}}, \bibinfo
		{author} {\bibfnamefont {X.-M.}\ \bibnamefont {Lu}},\ and\ \bibinfo {author}
		{\bibfnamefont {X.}~\bibnamefont {Wang}},\ }\href
	{https://doi.org/10.1088/1751-8121/ab5d4d} {\bibfield  {journal} {\bibinfo
			{journal} {J. Phys. A: Math. Theor.}\ }\textbf {\bibinfo {volume} {53}},\
		\bibinfo {pages} {023001} (\bibinfo {year} {2019})}\BibitemShut {NoStop}%
	\bibitem [{\citenamefont {Mera}\ \emph {et~al.}(2022)\citenamefont {Mera},
		\citenamefont {Zhang},\ and\ \citenamefont {Goldman}}]{SciPostMera2022}%
	\BibitemOpen
	\bibfield  {author} {\bibinfo {author} {\bibfnamefont {B.}~\bibnamefont
			{Mera}}, \bibinfo {author} {\bibfnamefont {A.}~\bibnamefont {Zhang}},\ and\
		\bibinfo {author} {\bibfnamefont {N.}~\bibnamefont {Goldman}},\ }\href
	{https://doi.org/10.21468/SciPostPhys.12.1.018} {\bibfield  {journal}
		{\bibinfo  {journal} {SciPost Phys.}\ }\textbf {\bibinfo {volume} {12}},\
		\bibinfo {pages} {018} (\bibinfo {year} {2022})}\BibitemShut {NoStop}%
	\bibitem [{\citenamefont {Yu}\ \emph {et~al.}(2024)\citenamefont {Yu},
		\citenamefont {Li}, \citenamefont {Chu}, \citenamefont {Mera}, \citenamefont
		{{\"U}nal}, \citenamefont {Yang}, \citenamefont {Liu}, \citenamefont
		{Goldman},\ and\ \citenamefont {Cai}}]{Yu_2024}%
	\BibitemOpen
	\bibfield  {author} {\bibinfo {author} {\bibfnamefont {M.}~\bibnamefont
			{Yu}}, \bibinfo {author} {\bibfnamefont {X.}~\bibnamefont {Li}}, \bibinfo
		{author} {\bibfnamefont {Y.}~\bibnamefont {Chu}}, \bibinfo {author}
		{\bibfnamefont {B.}~\bibnamefont {Mera}}, \bibinfo {author} {\bibfnamefont
			{F.~N.}\ \bibnamefont {{\"U}nal}}, \bibinfo {author} {\bibfnamefont
			{P.}~\bibnamefont {Yang}}, \bibinfo {author} {\bibfnamefont {Y.}~\bibnamefont
			{Liu}}, \bibinfo {author} {\bibfnamefont {N.}~\bibnamefont {Goldman}},\ and\
		\bibinfo {author} {\bibfnamefont {J.}~\bibnamefont {Cai}},\ }\href
	{https://doi.org/10.1093/nsr/nwae065} {\bibfield  {journal} {\bibinfo
			{journal} {Nat. Sci. Rev.}\ }\textbf {\bibinfo {volume} {11}},\ \bibinfo
		{pages} {nwae065} (\bibinfo {year} {2024})}\BibitemShut {NoStop}%
	\bibitem [{\citenamefont {Jankowski}\ \emph {et~al.}(2023)\citenamefont
		{Jankowski}, \citenamefont {Morris}, \citenamefont {Bouhon}, \citenamefont
		{{\"U}nal},\ and\ \citenamefont {Slager}}]{jankowski2023optical}%
	\BibitemOpen
	\bibfield  {author} {\bibinfo {author} {\bibfnamefont {W.~J.}\ \bibnamefont
			{Jankowski}}, \bibinfo {author} {\bibfnamefont {A.~S.}\ \bibnamefont
			{Morris}}, \bibinfo {author} {\bibfnamefont {A.}~\bibnamefont {Bouhon}},
		\bibinfo {author} {\bibfnamefont {F.~N.}\ \bibnamefont {{\"U}nal}},\ and\
		\bibinfo {author} {\bibfnamefont {R.-J.}\ \bibnamefont {Slager}},\ }\href
	{https://doi.org/10.48550/arXiv.2311.07545} {\bibfield  {journal} {\bibinfo
			{journal} {arXiv:2311.07545}\ } (\bibinfo {year} {2023})}\BibitemShut
	{NoStop}%
	\bibitem [{\citenamefont {Niu}\ \emph {et~al.}(1985)\citenamefont {Niu},
		\citenamefont {Thouless},\ and\ \citenamefont {Wu}}]{PhysRevB.31.3372}%
	\BibitemOpen
	\bibfield  {author} {\bibinfo {author} {\bibfnamefont {Q.}~\bibnamefont
			{Niu}}, \bibinfo {author} {\bibfnamefont {D.~J.}\ \bibnamefont {Thouless}},\
		and\ \bibinfo {author} {\bibfnamefont {Y.-S.}\ \bibnamefont {Wu}},\ }\href
	{https://doi.org/10.1103/PhysRevB.31.3372} {\bibfield  {journal} {\bibinfo
			{journal} {Phys. Rev. B}\ }\textbf {\bibinfo {volume} {31}},\ \bibinfo
		{pages} {3372} (\bibinfo {year} {1985})}\BibitemShut {NoStop}%
\end{thebibliography}
\end{document}